\documentclass[lettersize,journal]{IEEEtran}
\usepackage[noadjust]{cite}
\usepackage{amsmath,amssymb}
\usepackage{graphicx}
\usepackage{braket}
\usepackage{epsfig}
\usepackage{epstopdf}
\usepackage{mdwmath}
\usepackage{xcolor}
\usepackage{varwidth}
\usepackage{framed}
\usepackage{float}
\usepackage{placeins}
\usepackage{afterpage}
\usepackage{ragged2e}
\usepackage{amssymb}
\usepackage{pifont}

\usepackage{blindtext}
\usepackage{enumitem}
\usepackage{xcolor}
\usepackage{eqparbox}
\usepackage{fixltx2e}
\usepackage{multirow}
\usepackage{enumitem,lipsum}
\usepackage[utf8]{inputenc}
\usepackage[english]{babel}
\newcolumntype{P}[1]{>{\centering\arraybackslash}p{#1}}
\usepackage{graphicx}
\usepackage{amsmath, amssymb}
\usepackage{graphicx}
\usepackage{epsfig}
\usepackage{epstopdf}
\usepackage{array}
\usepackage{graphicx}
\usepackage{subcaption}
\usepackage{hyphenat}
\usepackage[english]{babel}
\usepackage[utf8]{inputenc}
\usepackage{algorithm}
\usepackage{physics}
\usepackage[noend]{algpseudocode}
\usepackage{cuted}
\usepackage{enumitem}

\newtheorem{theorem}{Theorem}[section]

\newtheorem{definition}{Definition}

\DeclareMathOperator*{\argmin}{arg\,min}

\usepackage{booktabs} 
\usepackage{graphicx}
\usepackage{caption}
\usepackage{bbding}
\usepackage{pifont}
\usepackage{wasysym}
\usepackage{amssymb}

\newtheorem{remark}{Remark}
\usepackage{framed,url,caption,graphicx}
\usepackage{amsmath,xcolor,soul,algpseudocode,algorithm,xspace,tabularx,multirow}
\usepackage{amssymb, cleveref}

\PassOptionsToPackage{hyphens}{url}

\newcommand{\squishenum}{   \begin{enumerate}{}    { \setlength{\itemsep}{0pt}      \setlength{\parsep}{0pt}      \setlength{\topsep}{3pt}       \setlength{\partopsep}{0pt}      \setlength{\leftmargin}{1.5em} \setlength{\labelwidth}{1em}      \setlength{\labelsep}{0.5em} } }
\newcommand{\squishlist}{   \begin{list}{$\bullet$}    { \setlength{\itemsep}{0pt}      \setlength{\parsep}{3pt}      \setlength{\topsep}{3pt}       \setlength{\partopsep}{0pt}      \setlength{\leftmargin}{1.5em} \setlength{\labelwidth}{1em}      \setlength{\labelsep}{0.5em} } }
\newcommand{\squishlisttwo}{   \begin{list}{$\bullet$}    { \setlength{\itemsep}{0pt}    \setlength{\parsep}{0pt}      \setlength{\topsep}{0pt}     \setlength{\partopsep}{0pt}      \setlength{\leftmargin}{2em} \setlength{\labelwidth}{1.5em}      \setlength{\labelsep}{0.5em} } }
\newcommand{\squishend}{    \end{list}  }
\newcommand{\squishenumend}{	\end{enumerate}	}
\usepackage[utf8]{inputenc}
\usepackage[english]{babel}
\usepackage{fancyhdr}

\usepackage{parselines} 
\usepackage{blindtext}

\usepackage{xcolor}
\usepackage{mathtools}
\newcommand{\cR}{\mathcal{R}}
\newcommand{\cL}{L}

\newcommand{\eat}[1]{}
\newcommand{\dqnum}{QPrimalDual}
\newcommand{\dqnumapproax}{QPrimalDual-approx}
\newcommand{\dqnumpatch}{QPrimalDual-DA}
\newcommand{\dqnumpi}{QPrimalDual-PI}
\newcommand{\dqnumpatchapprox}{QPrimalDual-DA-approx}

\title{A Framework for Distributed Resource Allocation in Quantum Networks
\thanks{This research is supported in part by the NSF grant CNS-2402861, NSF- ERC Center for Quantum Networks grant EEC-1941583. It is also sponsored in part by PiQSci project of Advanced Scientific Computing Research program, U.S. Department of Energy, and is performed at Oak Ridge National Laboratory managed by UT-Battelle, LLC for U.S. Department of Energy under Contract No. DE-AC05-00OR22725. This manuscript has been co-authored by UT-Battelle, LLC, under contract DE-AC05-00OR22725 with the US Department of Energy (DOE). The US government retains and the publisher, by accepting the article for publication, acknowledges that the US government retains a nonexclusive, paid-up, irrevocable, worldwide license to publish or reproduce the published form of this manuscript, or allow others to do so, for US government purposes. DOE will provide public access to these results of federally sponsored research in accordance with the DOE Public Access Plan (http://energy.gov/downloads/doe-public-access-plan).}
}

\author{Nitish K. Panigrahy*, Leonardo Bacciottini*, C. V. Hollot, Emily A. Van Milligen, Matheus Guedes de Andrade, Nageswara S. V. Rao, Gayane Vardoyan, Don Towsley
\thanks{* Authors with equal contribution}
\thanks{N. K. Panigrahy (npanigrahy@binghamton.edu) is with the Faculty of Binghamton University.}
\thanks{L. Bacciottini (lbacciottini@umass.edu), C. V. Hollot, M. G. de Andrade, G. Vardoyan and D. Towsley are with the University of Massachusetts, Amherst.}
\thanks{E. A. Van Milligen is with Virginia Tech.}
\thanks{Nageswara S. V. Rao is with Oak Ridge National Laboratory.}
}
\begin{document}
\maketitle
\begin{abstract}
We introduce a distributed resource allocation framework for the Quantum Internet that relies on feedback-based, fully decentralized coordination to serve multiple co-existing applications. We develop quantum network control algorithms under the mathematical framework of Quantum Network Utility Maximization (QNUM), where utility functions quantify network performance by mapping entanglement rate and quality into a joint optimization objective. We then introduce QPrimalDual, a decentralized, scalable algorithm that solves QNUM by strategically placing network controllers that operate using local state information and limited classical message exchange. We prove global asymptotic stability for concave, separable utility functions, and provide sufficient conditions for local stability for broader non-concave cases. To reduce control overhead and account for quantum memory decoherence, we also propose schemes that locally approximate global quantities and prevent congestion in the network.
We evaluate the performance of our approach via simulations in realistic quantum network architectures. Results show that QPrimalDual significantly outperforms baseline allocation strategies, scales with network size, and is robust to latency and decoherence. Our observations suggest that QPrimalDual could be a practical, high-performance foundation for fully distributed resource allocation in quantum networks.

\end{abstract}
\section{Introduction}

A full-fledged Quantum Internet demands meticulous orchestration between hardware components that must seamlessly cooperate to service a large number of end nodes deploying a myriad of applications. To realize this vision, many lessons can be learned from its classical counterpart. A vast network of networks, the Internet operates in a distributed manner wherein geographically distributed nodes communicate by sending data packets at rates dictated by data transport protocols like the Transmission Control Protocol (TCP) \cite{jacobson1988congestion}. Built-in congestion control mechanisms relay signals to data sources, instructing them to adjust sending rates to enable congested network segments to stabilize. These elegant feedback loops can be tailored to accommodate different quality-of-service (QoS) goals, such as high throughput, low latency, and fairness. Importantly, no single network node need be cognizant of the full structure of this global network, and no centralized control mechanism is necessary (nor practical), for all of this to work.

 With these principles in mind, we propose here a foundational framework for a decentralized Quantum Internet, focusing specifically on the machinery responsible for allocating network resources and providing end-to-end connectivity. 
Like the Internet, we envision that the Quantum Internet will also use carefully designed feedback mechanisms to achieve fully distributed operation. Our focus here will be on first- and second-generation quantum networks \cite{munro2015inside}, whose goal is to reliably distribute entanglement to end nodes. The concepts introduced in this work can also be adapted to third-generation networks, but this carries additional considerations that place this topic outside of our scope. While developing this distributed Quantum Internet framework, we take into consideration several obstacles unique to quantum networks:

\begin{itemize}[noitemsep,topsep=0pt,leftmargin=*]
    \item Hardware limitations, likely to be severe in the near-term, impose additional requirements on quantum entanglement distribution protocols that are not found in the classical setting. In the latter, virtually perfect storage, for instance, can be taken for granted. We propose algorithms that are quantum decoherence-aware;
    \item Quantum applications are highly sensitive to the quality of states that are provided to them. Well-designed performance measures for entanglement-based applications thus incorporate both entanglement quality as well as entanglement distribution rate. Our proposed algorithms are compatible with optimizing such performance measures;
    \item Classical messages are necessary to notify network nodes of the status of certain quantum communication subroutines (e.g., entanglement swapping \cite{zukowski1993event}, distillation outcomes \cite{dur2007entanglement}), but also to relay control information. This results in a higher amount of compulsory orchestration than seen in conventional networks.
\end{itemize}

At its core, our distributed Quantum Internet framework uses the mathematical tool known as Quantum Network Utility Maximization (QNUM), introduced by Vardoyan et al. in \cite{Vardoyan23QNUM}. There, the authors introduce the notion of a \textit{quantum utility function} -- a function that can be either prescribed or ascribed to a quantum application to capture the nature and amount of utility that a quantum user or application draws from the network. Quantum utility functions can be designed specific to each application, as well as provide fairness and stability guarantees. QNUM aims to maximize the aggregate utility across all users/applications actively using network resources (e.g., quantum channels, quantum memories at intermediate network nodes such as quantum switches and routers). In \cite{Vardoyan23QNUM}, Vardoyan et al. demonstrated the potential of QNUM to “tune” quantum networks, and provide satisfactory, fair service to all active users. However, they did not outline a method of solving the QNUM optimization problem in real-time, nor did they provide a practical algorithm to deploy the solution onto the network. As we will see, accomplishing these tasks warrants a separate, in-depth study which is the focus of our work.

Here, we use the QNUM framework as a foundation from which to derive our distributed entanglement distribution algorithms. This approach parallels that of classical NUM, introduced by Kelly in \cite{kelly1998rate, kelly1997charging} for conventional networks. 
NUM has been widely applied in classical settings -- see e.g., \cite{pham2016network, lee2006optimal,meshkova2010utility,xing2016utility} -- and notably, has provided deep insights into the design and behavior of different internet protocols, including the reverse- and forward-engineering of TCP \cite{kelly1998rate, kunniyur2003end, low2003duality, jin2004fast,lee2006optimal}. The QNUM framework could play a similar foundational role in the development of the Quantum Internet.

We make the following contributions:
\begin{itemize}[noitemsep,topsep=0pt,leftmargin=*]
\item 
We propose a distributed algorithm for solving the QNUM optimization problem. Our design features two types of interacting controllers that operate using local information and limited classical message exchange. We also introduce protocol variants that mitigate quantum memory decoherence and reduce explicit control message exchanges; 
\item We establish theoretical guarantees for the stability\footnote{In the context of distributed systems, stability refers to the system’s ability to return to equilibrium state (optimal resource allocation) following small (local) or any (global) perturbations and is critical to prevent undamped oscillations, which can severely degrade network performance \cite{jacobson1988congestion}.} of our proposed algorithm under various convexity conditions. Namely, we provide conditions for global asymptotic stability for concave utility functions, and for local asymptotic stability for non-concave functions;

\item We outline a practical implementation of our algorithms in a \emph{sequential quantum network} -- one in which entanglement swapping is performed sequentially hop-by-hop at intermediate nodes along a path connecting end nodes. Our simulation results demonstrate that the proposed algorithms substantially outperform baseline resource allocation strategies and remain robust against latency and quantum memory decoherence.
\end{itemize}

The remainder of this paper is organized as follows. In Section~\ref{sec:model}, we define the network model and discuss the QNUM formulation. We present our distributed algorithm in Section~\ref{sec:dqnum} and outline a protocol for implementing it in sequential quantum networks in Section~\ref{sec:sequential}. We evaluate the performance of our algorithm and variants through simulated experiments in Section~\ref{sec:perf-eval}. We review related work in Section~\ref{sec:rel}. Finally, we conclude the paper in Section~\ref{sec:dis}. 
\section{Quantum Network Model}\label{sec:model}
We present the quantum network model used in this work, focusing on bipartite entanglement distribution between pairs of network users. While our schemes can be applied to multipartite entanglement, this extension requires a careful study which we leave for the future.

We model a quantum network as a graph $G=(V, \cL)$, where $V$ denotes the set of vertices/nodes and $\cL$ is the set of physical links (e.g., optical fiber) representing quantum communication channels between neighboring nodes. We assume that each link $l \in \cL$ generates heralded bipartite entanglement between its adjacent nodes using a single photon entanglement generation scheme \cite{cabrillo1999creation, humphreys2018deterministic}. The entanglement generation process is parametrized by a tunable brightness parameter $\alpha_l$, which controls the fidelity and rate of generated entanglement. In particular, the generated state is of the form
\begin{align}
    \rho_l = (1-\alpha_l)\ketbra{\Psi^+} + \alpha_l \ketbra{\uparrow\uparrow}, \label{eq:brightstate}
\end{align}
\noindent where $\ket{\Psi^+}$ is the desired entangled Bell state of the form $(\ket{\uparrow\downarrow}+\ket{\downarrow\uparrow})/\sqrt{2}$ and $\ket{\uparrow\uparrow}$ is the bright state. The state $\rho_l$ is generated between the nodes of link $l$ with probability $2\eta_l\alpha_l$. Here, $\eta_l$ scales as $e^{-\beta(L_l/2)}$, where $L_l$ is the physical length of link $l$ and $\beta$ is the attenuation coefficient. We model the resulting link level entangled state (LLE) as a Werner state (up to a unitary rotation), even though the actual quantum state is of the form \eqref{eq:brightstate}. I.e., mathematically, we approximate $\rho_l$ as
\begin{align}
    \rho_l = w_l\ketbra{\Psi^+} + (1-w_l)\frac{\mathbb{I}}{4},
    \label{eq:werner_state}
\end{align}
where the associated Werner parameter $w_l \in [0,1]$ is a linear function of $\alpha_l$, and is chosen so that the fidelity of the state \eqref{eq:brightstate} matches that of \eqref{eq:werner_state}. In effect, this is a twirling operation \cite{bennett1996mixed} applied to the state in \eqref{eq:brightstate}. Twirling is commonly used to simplify analysis, although we do not assess here its effects on overall performance. The rate at which $\rho_l$ is generated, which we refer to as the \textit{capacity} of link $l$, can be expressed as \cite{Vardoyan23QNUM}:
\begin{align}
    c_l = \chi_l\eta_l\alpha_l = 3\chi_l\eta_l(1-w_l)/2 = d_l(1-w_l),\label{eq:capacity}
\end{align}
where $\chi_l$ represents the repetition rate (or attempt rate) of the entanglement generation process and $d_l =3\chi_l\eta_l/2.$ The fidelity of $\rho_l$ is given by $F_l = (3w_l+1)/4.$ As $w_l$ increases, so does the LLE fidelity, while the term $(1-w_l)$ decreases, reducing the capacity of the link. This captures the fundamental rate-fidelity trade-off of the single-click scheme: generating higher-fidelity LLE reduces the throughput a link can support.

When a pair of end nodes $(A,B)$ wishes to share end-to-end (E2E) entanglement, the end nodes initiate an \textit{entanglement session}, which is characterized by $A$ and $B$'s requested entanglement features such as average successful generation rate and fidelity.
We denote by $\cR$ the set of all entanglement sessions, where each session $r \in\cR$ corresponds to a node-pair $(A_r, B_r)$ to whom the session belongs. A session is initiated by first identifying a path in $G$ that connects  $A_r$ and $B_r$. Entanglement swapping is then performed at intermediate network nodes by fusing the LLEs along this path to generate E2E states between $A_r$ and $B_r$. For simplicity, we assume that entanglement swap operations are deterministic, even though our framework can accommodate probabilistic swaps. We also assume that the path connecting $(A_r, B_r)$ is predetermined and known in advance. For notation simplicity, we will also use the variable $r$ to denote the path associated with session $r$, as is standard in classical NUM literature. Note that multiple sessions between a node-pair can be modeled by creating auxiliary nodes with perfect internal connectivity.

Let $R_r$ and $F_r$ denote the rate and fidelity of the average E2E entangled state served to the end nodes of session $r$. In other words, $A_r$ and $B_r$ receive a Werner state with fidelity $F_r$ at a rate of $R_r$ states per second. When each LLE is modeled as a Werner state, the resulting E2E entangled state after entanglement swapping along the path also remains a Werner state. The corresponding Werner parameter is given by $W_r = \prod_{l\in r}w_l.$ Thus, the fidelity of the generated E2E entangled state is
$F_r = 3/4\prod_{l\in r}w_l+1/4.\nonumber$ 
From this equation, it is clear that entanglement swapping with sub-unit $w_l$'s results in a post-swap state with lower fidelity.
When quantum memory decoherence is taken into account, the fidelity further degrades over time. While the QNUM formulation does not account for memory decoherence, we incorporate and mitigate its effects later in Section \ref{subsec:ext}. Mechanisms to combat gate noise can similarly be incorporated into our algorithms, although we leave this for future work.

\begin{figure*}[!htbp]
\centering
\begin{minipage}{0.4\textwidth}
\includegraphics[width=1\textwidth]{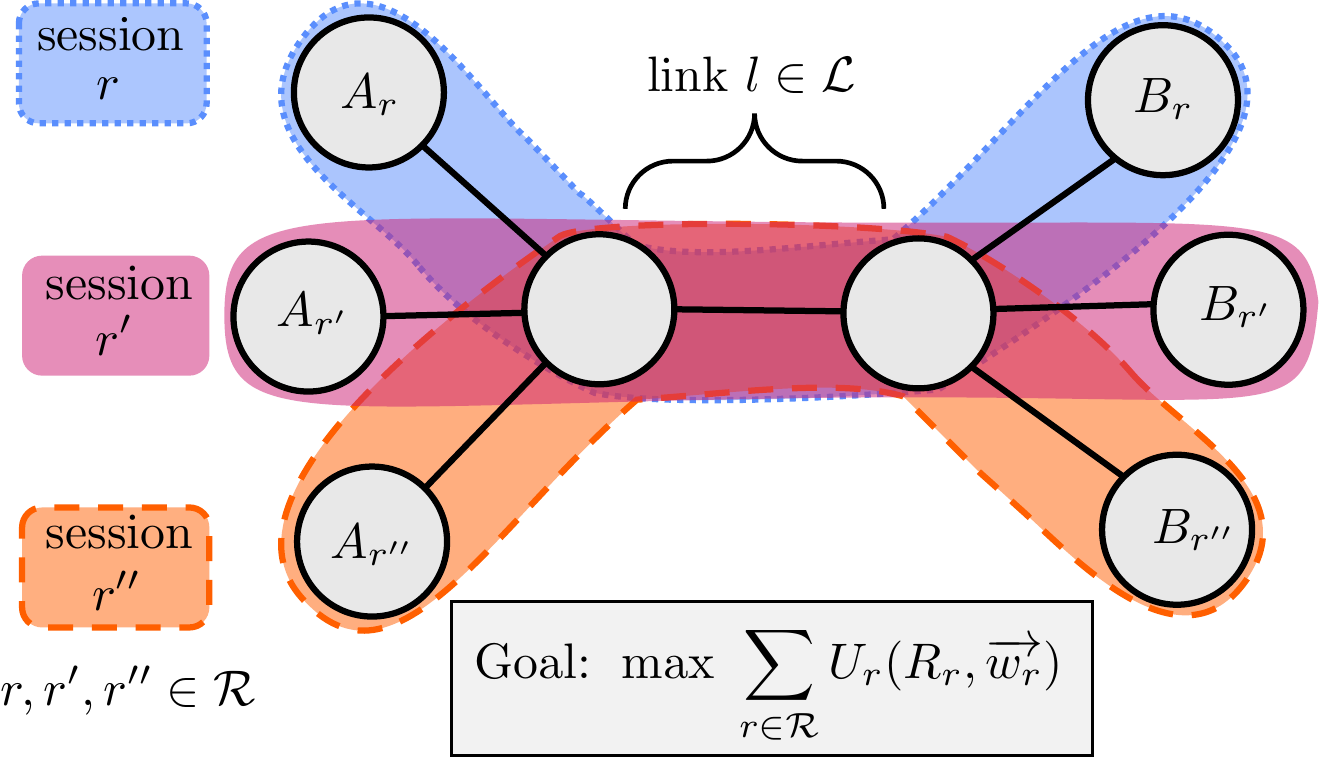}
\subcaption{}\label{fig:qnum-a}
\end{minipage}
\hspace{0.2cm}
\begin{minipage}{0.5\textwidth}
\includegraphics[width=1\textwidth]{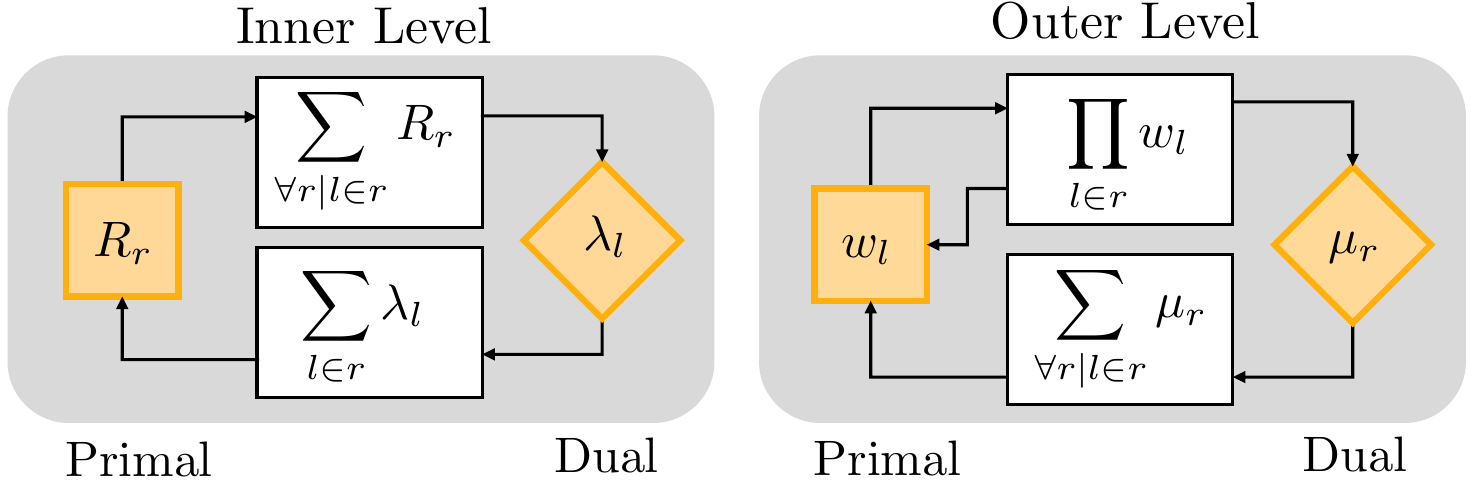}
\subcaption{}\label{fig:qnum-b}
\end{minipage}
\caption{Quantum Network Utility Maximization and primal-dual solution. (a) Three sessions ($r, r', r''$) need to create bipartite entanglement between corresponding pairs of end nodes.  The goal is to maximize the aggregate utility across all sessions. (b) A distributed solution via primal-dual approach: In each iteration, it updates two sets of primal variables ($R_r, w_l$) and dual variables ($\lambda_l, \mu_r$), based on feedback received from both link and session controllers. The details of the inner and outer-level update structure are discussed in Section \ref{sub:two-level}.}
\label{fig:qnum}
\end{figure*}

We assume that each pair of end nodes in session $r$ is associated with a utility function $U_r(R_r,F_r)$, where $U_r$ is a differentiable function on $(R_r,F_r)$ and the fidelity $F_r$ is determined by the vector $\vec{w}_r$ of elementary link Werner parameters along path $r$. Let $f_r(R_r,\vec{w}_r) \equiv \partial U_r(R_r,F_r)/\partial R_r$. We assume that $f_r(R_r,\vec{w}_r)$ is monotone in $R_r.$ The QNUM problem \cite{Vardoyan23QNUM} (Figure \ref{fig:qnum-a}) aims to configure link-level Werner parameters $w_l$ and session-level E2E rates $R_r$ that maximize the aggregate utility, $\sum_{r\in \cR}  U_r(R_r,F_r),$ while satisfying QoS
and link capacity constraints. Formally, QNUM is defined as
\begin{align} 
\textbf{QNUM: }\max & \quad \sum_{r\in \cR}  U_r(R_r,\vec{w}_r) 
\label{eq:optimal_qnum_obj}
\\
\text{subject to}\quad
& \nonumber \\
\quad & \sum_{r:l\in r}R_r  \leq  d_l(1-w_l), \quad \forall{l \in \cL},\label{eq:optimal_qnum_cap} \\
\quad & \sum_{l:l\in r} \log w_l \ge K_r, \quad \forall{r \in \cR},\label{eq:optimal_qnum_fid}\\
\quad & 0 \leq  w_l \leq 1, \quad \forall{l \in \cL},\label{nonnegativity-w}\\
\quad & R_r \geq  0, \quad \forall{r \in \cR}.\label{nonnegativity-R}
\end{align} 
\noindent The capacity of link $l$ for the single photon scheme is expressed as $d_l(1-w_l),$ where $d_l$ is a link specific parameter and serves as an input to the optimization problem (see \eqref{eq:capacity}). Constraint \eqref{eq:optimal_qnum_cap} ensures that the aggregate E2E rate of sessions using a given link do not exceed the link capacity. Constraint \eqref{eq:optimal_qnum_fid} imposes minimum thresholds on E2E entanglement that may not be directly accounted for within $U_r(R_r,\vec{w}_r)$. Specifically, certain utility functions inherently incorporate a minimum fidelity threshold $F_r^{\text{min}}$ on the E2E entangled state, since any fidelity below the threshold simply yields a utility of zero. The secret key rate (SKR) of entanglement-based BB84 QKD protocol is one example of this -- when implemented with Werner states, it is given by
\begin{align}\label{eq:skr_raw}
     \text{SKR}_{\text{BB84}} =  \max\left\{0, R_r \bigg( 1 - 2 h\bigg( \frac{1}{2} - \frac{1}{2}\prod\limits_{l:l\in r}w_l\bigg)\bigg)\right\},
\end{align}
which we express in terms of the individual LLE Werner parameters, and with $h$ the binary entropy function \cite{bennett2014quantum, lo2005efficient, shor2000simple, PhysRevLett.68.557, murta2020key}.
In the most general case, however, we permit arbitrary user-, application-, or network operator-defined utility functions that might not share this feature; in such cases, it may be more convenient to have an explicit constraint such that $F_r\geq F_r^{\text{min}}$. To keep the notation within the QNUM problem \eqref{eq:optimal_qnum_obj}-\eqref{nonnegativity-R} consistently in terms of Werner parameters, we express these constraints in terms of $w_l$'s. Namely, in \eqref{eq:optimal_qnum_fid}, ${K_r \coloneq \log[(4F_r^{\text{min}}-1)/3]}$ encapsulates session $r$'s E2E Werner parameter, where $F_r^{\text{min}}$ is the minimum fidelity threshold specified by session $r$'s application. Adherence to \eqref{eq:optimal_qnum_fid} is then equivalent to ensuring $F_r\geq F_r^{\text{min}}$.

To ensure fairness across sessions, we follow the approach in \cite{Vardoyan23QNUM} and take the logarithm of $\text{SKR}_{\text{BB84}}$, resulting in the following utility function:
\begin{align}
        U_r^\text{SKR}(R_r,\vec{w}_r) = 
                        \log\bigg( R_r\bigg( 1 - 2 h\bigg( \frac{1}{2} - \frac{1}{2}\prod\limits_{l:l\in r}w_l\bigg)\bigg)\bigg). \label{eq:skr}  
\end{align}
Note that the $\max\{0,\cdot\}$ operator in $\text{SKR}_{\text{BB84}}$ in \eqref{eq:skr_raw} ensures non-negativity. In $U_r^\text{SKR},$ the non-negativity of the argument inside the logarithm in \eqref{eq:skr} can be ensured by setting appropriate values of $K_r$ in constraint \eqref{eq:optimal_qnum_fid}.

We also consider a utility function based on entanglement negativity \cite{vidal2002computable}. This is an easy to compute measure of entanglement which was proposed for mixed states of arbitrary bipartite systems. For Werner states, the entanglement negativity-based utility function is defined as \cite{Vardoyan23QNUM}
\begin{align}
    U_r^\text{neg}(R_r,\vec{w}_r) =  \log\bigg( R_r \bigg(3\prod_{l:l\in r}w_l-1\bigg)\bigg).\label{eq:neg}
\end{align}

The QNUM optimization problem defined in \eqref{eq:optimal_qnum_obj}-\eqref{nonnegativity-R} can be solved by a centralized controller that requires complete access to network topology, path information, and both session and link-level data for all sessions and links. Additionally, once the optimal solution is computed, it must be made available to all links and sessions with a sufficiently small delay to avoid deployment of the solution on a completely outdated network state. However, in large and/or dynamic networks where link conditions, entanglement sessions, and even topology change frequently,  centralized computation and deployment of the solution are intractable. 
\section{A Distributed Algorithm for Resource Allocation}\label{sec:dqnum}
Here, we present our resource allocation algorithm, which operates in a fully decentralized manner much like its Internet analogues, such as the widely-adopted TCP congestion control algorithm \cite{jacobson1988congestion}.
This algorithm is a direct result of a distributed solution to the QNUM problem \eqref{eq:optimal_qnum_obj}-\eqref{nonnegativity-R}, which we obtain via a \textit{primal-dual} optimization approach commonly applied in optimization and classical network control theory \cite{srikant2004book,srikant2014communication}.
We also provide guidelines for efficiently deploying our algorithm. At a high level, its implementation requires two types of controllers: one set situated at network end nodes, and the other set at quantum links. Together, these controllers gather and feed information to each other, which they then use to dynamically adjust quantum network parameters, such as LLE generation rates and fidelities, to reflect and adapt to current network conditions.
\subsection{Distributed solution to QNUM via primal-dual approach}\label{sub:dqnumPD}
In this algorithm, which we refer to as \dqnum, we introduce a {\it dual} variable $\lambda_l$ associated with capacity constraint \eqref{eq:optimal_qnum_cap} on link $l$. Similarly, another dual variable $\mu_r$, is introduced corresponding to the minimum fidelity constraint \eqref{eq:optimal_qnum_fid} for session $r$. We refer to the optimization variables that represent resource allocation decisions ($R_r$'s and $w_l$'s) as {\it primal} variables. To simplify notation and analysis, we introduce vector representations of the primal and dual variables, i.e. $\vec{R}$, $\vec{w}$, $\vec{\lambda}$ and $\vec{\mu}$ denote the vectors of all session rates, link-level Werner parameters, link-level and session-level dual variables, respectively. The Lagrangian function is given by
\begin{align}\label{eq:lag}
\mathcal{A} (\vec{R}, \vec{w}, \vec{\lambda}, \vec{\mu}) \coloneq &\sum_{\substack{{r\in \cR}}}  U_r(R_r,\vec{w}_r) \nonumber\\
&- \sum_{l}\lambda_l\bigg[\sum_{r:l\in r}R_r-d_l(1-w_l)\bigg] \nonumber\\
&- \sum_{r}\mu_r\bigg[K_r - \sum_{l:l\in r}\log w_l\bigg]. 
\end{align}
Note that the Lagrangian function of this problem includes three components: the original objective, a penalty term for violating the link capacity constraints and another penalty term  for violating the minimum fidelity constraints. We define the Lagrangian dual as
\begin{align}
\mathcal{D}(\vec{\lambda}, \vec{\mu}) = \max\limits_{\vec{R}\ge\vec{0}, \vec{0}\le\vec{w}\le\vec{1}} \mathcal{A} (\vec{R}, \vec{w}, \vec{\lambda}, \vec{\mu}).\label{eq:lag-dual}
\end{align}

A solution to QNUM is obtained by minimizing the dual function $\mathcal{D}$ over all non-negative $\vec{\lambda}$ and $\vec{\mu}$. To achieve this, we use the gradient method \cite{chiang2006decomposition} to iteratively update the primal and dual variables, where updates are calculated using the gradients of the Lagrangian function $\mathcal{A}$ (see updates I-IV presented below). The iterative updates continue until the Karush–Kuhn–Tucker (KKT) conditions  \cite{kuhn2013nonlinear} on $\mathcal{A}$ are satisfied. While maximizing $\mathcal{A}$ in \eqref{eq:lag-dual}, we assume that the optimal values of $R_r$ are non-negative. As an example, this holds for the SKR and entanglement negativity utility functions. Note also that the constraint $\vec{0}\le\vec{w}\le\vec{1}$ in \eqref{eq:lag-dual} follows implicitly since the expression $\log w_l$ in \eqref{eq:lag} is only defined for $w_l \ge 0.$ Similarly, when $R_r \ge 0$ for all $r\in\cR$, the minimum possible value of $\sum_{r:l\in r} R_{r} = 0.$ In such a case, setting $w_l > 1 $ would make the right-hand side of the link capacity constraint in \eqref{eq:optimal_qnum_cap} negative, making the constraint infeasible. Therefore, for every link, $w_l$ must satisfy $w_l \le 1.$

We define an iteration to be a discrete time step (indexed by $t$), during which \dqnum\;updates the primal and dual variables. The update rules of the \dqnum\;algorithm are as follows.

{\bf Notation:} Let $\lambda_l(t)$, $\mu_r(t)$, $w_l(t)$, $R_r(t)$ refer to the values of $\lambda_l$,  $\mu_r$, $w_l$, $R_r$ at iteration $t$.

{\bf Initialize:} $\lambda_l(0), w_l(0), R_r(0)$ and $\mu_r(0)$ to random feasible values, i.e., $\lambda_l(0), \mu_r(0) > 0$, $0\le w_l(0)\le 1$ and $R_r(0) > 0.$

{\bf Primal-dual Update:} Iteratively apply the following update rules for $t = 1,\dots$ until convergence. 
\begin{enumerate}[label=\Roman*.]
    \item Link price updates: update $\lambda_l(t)$ as follows:
    \begin{align}\label{dual-lambda}
    &\dot{\lambda_l}(t) = \frac{-\partial\mathcal{A} (\vec{R}, \vec{w}, \vec{\lambda}, \vec{\mu})}{\partial\lambda_l} = \left[\sum_{r:l\in r}R_r(t)-d_l(1-w_l(t))\right],\nonumber
    \\
    &\lambda_l(t+1)\leftarrow\max\left\{\lambda_l(t) + k_{\lambda_l}(t)\dot{\lambda_l}(t),0\right\}.
\end{align}
    Here $\lambda_l(t)$ can be interpreted as the price incurred for exceeding the link capacity.
    \item Session rate updates: Recall that $f_r(R_r,\vec{w}_r) \equiv \partial U_r(R_r,\vec{w}_r)/\partial R_r$. Taking the derivative of $\mathcal{A}$ with respect to $R_r$ and equating it to zero, we get
    \begin{align}
        f_r(R_r,\vec{w}_r) = \sum_{l:l\in r}\lambda_l.\nonumber
    \end{align}
    We assume that $f_r(R_r,\vec{w}_r)$ is monotone in $R_r.$ For example, the utility functions $U^{\text{SKR}}_r$ and $U^{\text{neg}}_r$ satisfy this condition. Then the inverse function $f_r^{-1}(\cdot, \vec{w}_r)$ exists and we set
    \begin{align}\label{dual-rate}
    R_r(t+1) \leftarrow f_r^{-1}\bigg(\sum_{l:l\in r}\lambda_l(t), \vec{w}_r(t)\bigg).
\end{align}
    \item E2E Fidelity price update: $\mu_r(t)$ can be interpreted as the price incurred for violating the minimum E2E fidelity constraint; it is updated as follows:
    \begin{align}\label{dual-mu}
    &\dot{\mu}_r(t) = \frac{-\partial\mathcal{A} (\vec{R}, \vec{w}, \vec{\lambda}, \vec{\mu})}{\partial\mu_r} = \left[K_r - \sum_{l:l\in r}\log w_l(t)\right],\nonumber\\
    &\mu_r(t+1)\leftarrow\max\left\{\mu_r(t) + k_{\mu_r}(t)\dot{\mu_r}(t),0\right\}.
\end{align}
    \item Link-level Werner parameter update: let $f_l(R_r,\vec{w}_r) \equiv \frac{\partial U_r(R_r,\vec{w}_r)}{\partial w_l}$. The $w_l$'s are updated as:
    \begin{align}\label{dual-w}
    &\dot{w}_l(t) = \frac{\partial\mathcal{A} (\vec{R}, \vec{w}, \vec{\lambda}, \vec{\mu})}{\partial w_l} \nonumber\\
    &\quad\quad= -d_l\lambda_l(t)+\sum\limits_{r:l\in r}f_l(R_r(t),\vec{w}_r(t)) + \frac{\sum\limits_{r:l\in r}\mu_r(t)}{w_l(t)},\nonumber\\
    &\bar{w}_l(t) = \max\left\{w_l(t) + k_{w_l}(t)\dot{w}_l(t),0\right\}, \nonumber\\
    &w_l(t+1)\leftarrow\min\{\bar{w}_l(t),1\}.
    \end{align}
\end{enumerate}
The strictly positive parameters $k_{\lambda_l}(t), k_{\mu_r}(t),$ and $k_{w_l}(t)$ in the primal-dual update stage are known as scaling parameters or step sizes. They affect the convergence behavior of the primal-dual algorithm in terms of speed and stability.

To implement the above primal-dual algorithm in a distributed way, we propose the following set of controllers:
\begin{itemize}
\item A {\em link controller} is placed at the LLE generation source of each link and executes the link price \eqref{dual-lambda} and link-level Werner parameter \eqref{dual-w} updates;

\item A {\em session controller} updates the session rate and E2E fidelity price according to  \eqref{dual-rate} and \eqref{dual-mu}, respectively.
\end{itemize}

Note that to perform the corresponding updates, both link and session controllers require access to local and global information (as summarized in Figure \ref{fig:qnum-b}). Thus, explicit message passing between controllers may be necessary. However, depending on the way in which a quantum network is operated, certain global information can be locally approximated or efficiently embedded within classical control messages that are already being transmitted. In Section \ref{sec:sequential}, we demonstrate techniques for locally approximating global information in quantum networks based on sequential entanglement swapping, reducing the need for extensive inter-controller communication.

\subsection{Stability analysis of primal-dual}\label{subsec:stability}
In feedback-based distributed systems, stability ensures that the system returns to equilibrium after being perturbed (say due to noise or delay). It prevents unbounded behavior, allowing the system to revert back to predictable behavior over time. This raises a question: {\it Is the equilibrium point of the \dqnum\;algorithm stable, i.e., do the primal and dual variables converge over time after perturbations?} Below, we define  important stability notions within the framework of our algorithm and then answer the above question.

Define $(\vec{R}^*, \vec{w}^*, \vec{\lambda}^*, \vec{\mu}^*)$ as the equilibrium point of the system of differential equations derived from the update rules \eqref{dual-lambda}-\eqref{dual-w}. These differential equations are provided in the Appendix, see~\eqref{eq: DE}.
\begin{definition}
The equilibrium point of the differential equations corresponding to  \dqnum\;is globally asymptotically stable if, for any initial values of the primal and dual variables, the system trajectories converge to equilibrium  as time progresses:
\begin{align}
\lim_{t \to \infty} (\vec{R}(t), \vec{w}(t), \vec{\mu}(t), \vec{\lambda}(t)) = (\vec{R}^*, \vec{w}^*, \vec{\lambda}^*, \vec{\mu}^*).
\end{align}
\end{definition}

\begin{definition}
The equilibrium point of the differential equations corresponding to \dqnum\;is locally asymptotically stable if, for initial values of the primal and dual variables in the close neighborhood ($\cal{N^*}$) of the equilibrium, the system trajectories remain close and converge to equilibrium as time progresses:
\begin{align}
\lim_{t \to \infty} (\vec{R}(t), \vec{w}(t), \vec{\mu}(t), \vec{\lambda}(t)) = (\vec{R}^*, \vec{w}^*, \vec{\lambda}^*,\vec{\mu}^*),  \\\nonumber
(\vec{R}(0), \vec{w}(0), \vec{\mu}(0), \vec{\lambda}(0)) \in \cal{N^*}.
\end{align}
\end{definition}

To analyze the stability of \dqnum, we assume the following.

{\bf Separability assumption:} For our stability analysis, we assume the that utility functions are differentiable and separable: $\partial U_r/\partial R_r$ is only a function of $R_r$ and $\partial U_r/\partial {w_l}$ is only a function of $\vec{w}_r$. The secret key rate  and the entanglement negativity utility functions defined in Section \ref{sec:model} have this separable property. 

{\bf No feedback delay assumption:} Additionally, we assume that the network provides instant feedback, i.e., the information exchange happens instantaneously, and the variables needed to compute \eqref{dual-lambda}-\eqref{dual-w} are instantly available to the controllers.

{\bf Existence of equilibrium:} We assume the existence of an equilibrium solution $(\vec{R}^*, \vec{w}^*, \vec{\lambda}^*, \vec{\mu}^*)$. When the utility functions are concave, QNUM becomes a convex optimization problem. Slater’s conditions for QNUM also hold (See Appendix \ref{sub:slater-appendix}). Thus, strong duality applies, i.e., the optimal value of the dual problem \eqref{eq:lag-dual}  is equal to the optimal value of the QNUM problem defined in \eqref{eq:optimal_qnum_obj}-\eqref{nonnegativity-R}. 

{\bf Full column rank assumption for the routing matrix:} Let the routing matrix {\bf R} be defined such that each entry ${\bf R}_{lr}$ equals to $1$ if session $r$ uses link $l$ and $0$ otherwise. For the stability results to hold, we assume that {\bf R} has full column rank. Physically, this assumption means that the set of paths taken by the sessions are sufficiently diverse so that no session’s path can be expressed as a linear combination of others\footnote{Redundant sessions (sessions that have the same path) or sessions whose paths are combinations of paths of other sessions (e.g., session $r_1$ uses link $l_1$, session $r_2$ uses link $l_2$ and session $r_3$ uses both link $l_1$ and link $l_2$) are a few examples where the full rank assumption is violated.}. In future work, we aim to relax this assumption. We expect that even without full rank, \dqnum\;will still converge to a unique primal solution $\vec{R}^*$ and $\vec{w}^*$ (since QNUM reduces to a convex optimization problem under concave utilities). However, in this case, multiple dual optimal solutions ($\vec{\lambda}^*$ and $\vec{\mu}^*$) may exist. Similar observations have been reported for dual-based formulations of the classical NUM problem~\cite{tang2007equilibrium}.

For concave and non-concave utilities, the following theorems show that the \dqnum\; algorithm is stable; i.e., the solution $\vec{R}$, $\vec{w}$, $\vec{\lambda}$ and $\vec{\mu}$ converge to their equilibrium values.
\begin{theorem}
{\bf Concave Utilities}: Assume that $U_r(R_r,\vec{w_r})$ is concave and separable over $(R_r,\overrightarrow{w_r}).$ Under this condition and other previously mentioned assumptions, the equilibrium point $(\vec{R}^*, \vec{w}^*, \vec{\lambda}^*, \vec{\mu}^*)$ is globally asymptotically stable.
\end{theorem}
\begin{theorem}
{\bf Non-concave Utilities}: If $U_r(R_r,\overrightarrow{w_r})$ is separable, but not necessarily concave 
over $(R_r,\overrightarrow{w_r}),$ and satisfies $U''_{w_\ell}(w_\ell^*) < \sum_{r:\ell \in r} \mu_r^*/w_{\ell}^{*2}$ at equilibrium point $(w_\ell^*,\mu_r^*)$, then the equilibrium point $(\vec{R}^*, \vec{w}^*, \vec{\lambda}^*, \vec{\mu}^*)$  is locally asymptotically stable under previously mentioned assumptions.
\end{theorem}

Detailed proofs of the stability results for these theorems can be found in the Appendix. In our proofs, we assume that the network, as captured by \eqref{dual-lambda}-\eqref{dual-w}, follows continuous-time dynamics. We show that for concave utility functions, these equations will converge to equilibrium as $t\to\infty$ by using a Lyapunov argument and Lasalle’s invariance principle \cite{editionnonlinear}. Analyzing the actual network as a discrete-time system and accounting for delays in information exchange into the stability analysis remain our future work.
\begin{figure*}[!htbp]
\centering
\begin{minipage}{0.52\textwidth}
\includegraphics[width=1\textwidth]{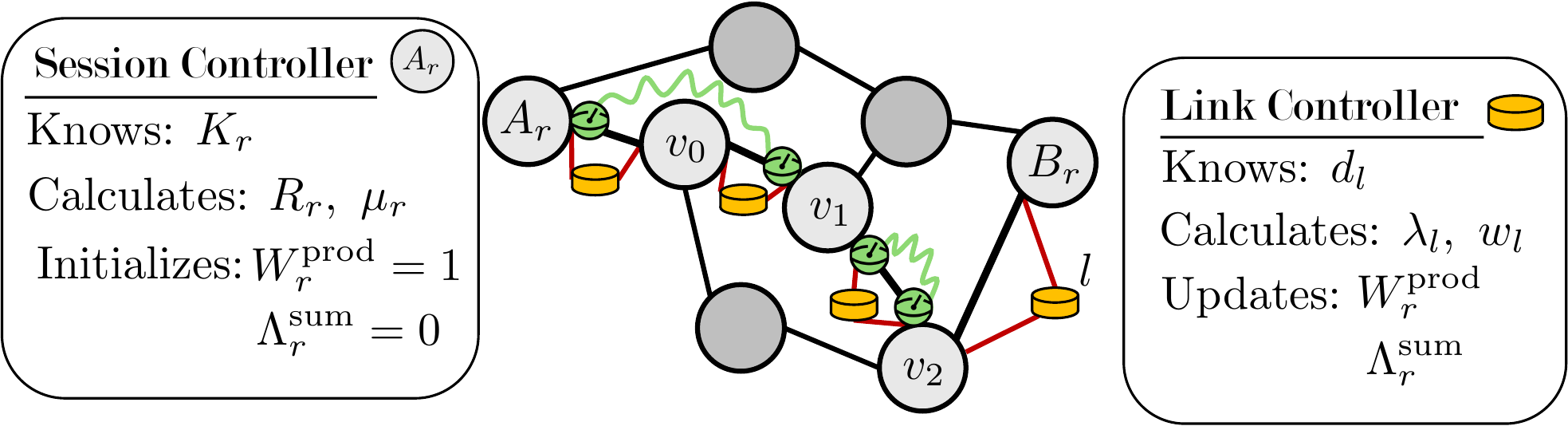}
\subcaption{}\label{fig:seq-qnum-a}
\end{minipage}
\hspace{0.1cm}
\begin{minipage}{0.45\textwidth}
\includegraphics[width=1\textwidth]{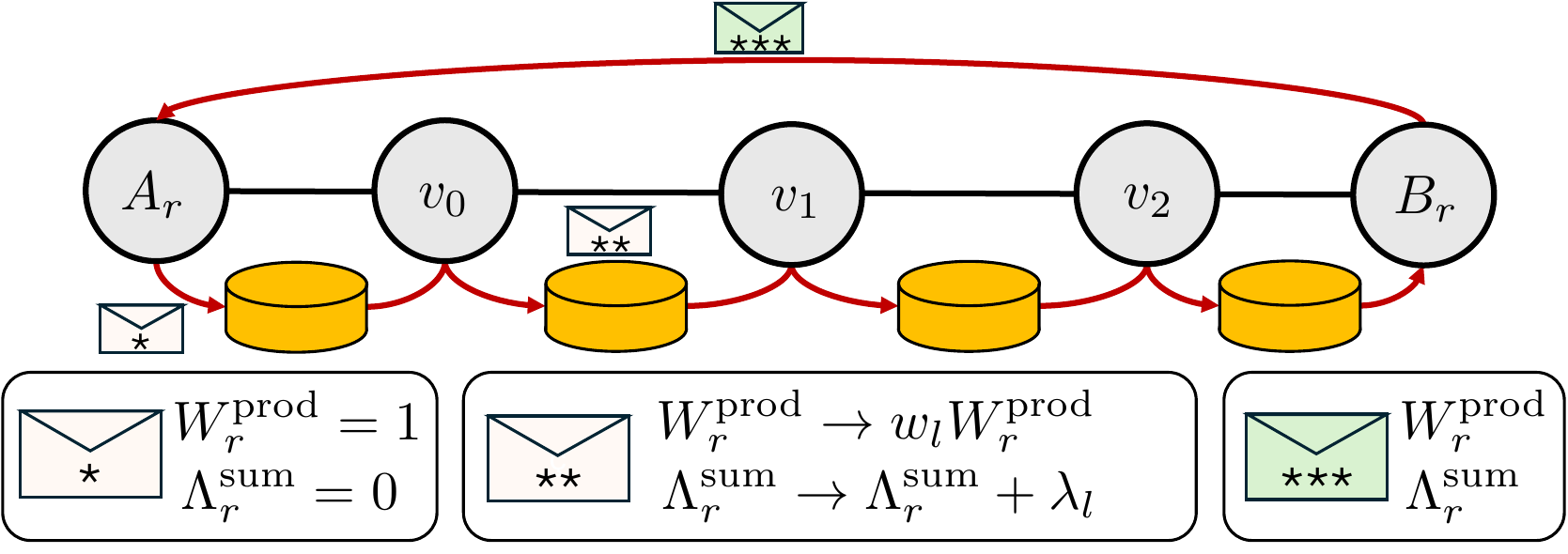}
\subcaption{}\label{fig:seq-qnum-b}
\end{minipage}
\caption{A summary of \dqnum\;implementation on a sequential quantum network architecture. (a) The information (primal and dual variables) stored and maintained by session and link controllers. (b) How the \dqnum\;information in the q-datagram header is updated along the way by link controllers, and relayed back through the ACK. Each link controller updates two header fields: $\Lambda_r^{\text{sum}}$ and $W_r^{\text{prod}}$ (shown as two-star mailbox symbol). The ACK, represented by a three-star mailbox symbol, relays back the final aggregated values of these two header fields.}
\label{fig:seq-qnum}
\end{figure*}
\subsection{An efficient bi-level implementation}\label{sub:two-level}
Among the different set of update rules, the link-level Werner parameter updates in \eqref{dual-w} require the greatest exchange of information between the remote controllers. To mitigate the issue of excessive delay due to message exchange, 
we propose a hierarchical bi-level update scheme. We adopt a separation of time scales, where the inner level  operates at a faster rate compared to the outer level (See Figure \ref{fig:qnum-b}). Similar bi-level schemes have been explored in classical wireless networking \cite{palomar2007alternative,johansson2005primal}.
\begin{itemize}[leftmargin=*]
    \item {\bf Inner Level:} At this level, we iteratively update the link prices and session rates using \eqref{dual-lambda} and \eqref{dual-rate} respectively.
    \item {\bf Outer Level:} For the outer level, we update the E2E fidelity prices and link-level Werner parameters via \eqref{dual-mu} and \eqref{dual-w} respectively once every $T_{\text{outer}}$ iterations, where $T_{\text{outer}}$ is the outer-loop period. The reduced update frequency at this level minimizes inter-controller communication.
\end{itemize}

Although theoretical stability guarantees provided in Section~\ref{subsec:stability} depend on the outer-level updates being executed at every iteration (i.e., $T_{\text{outer}} = 1$), our simulations (Section \ref{subsec:bilevel-robustness}) demonstrate that the algorithm continues to exhibit stability for higher values of the outer level update period.
\begin{remark}
    Note that, the stability analysis in Section~\ref{subsec:stability} is at the differential equation level, while the bi-level update scheme is an implementation level consideration. Sufficiently fast updates at the outer level will result in a stable system, given the existing theoretical guarantees. One can perform a more detailed analysis to provide explicit upper bounds on the outer level update rate necessary for ensuring  stability.
\end{remark}

\section{Implementation in Sequential Quantum Networks}\label{sec:sequential}
The \dqnum\;algorithm developed in Section \ref{sec:dqnum} provides theoretical guarantees based on continuous-time dynamics. However, its practical implementation requires consideration of discrete updates, inter-node propagation and queueing delays, and the resulting quantum memory decoherence. In this section, we focus on sequential quantum networks -- networks where entanglement swapping is performed sequentially  \cite{bacciottini2024leveraging, chenhu2025sequential}
-- and describe how \dqnum\;can be implemented within them. Sequential quantum networks have recently been shown to achieve performance on par with other swapping-based architectures in various settings \cite{pouryousef2024minimal}. Their reduced implementation overhead \cite{xiao2023connectionless, bacciottini2024leveraging} makes them  practical candidates for scalable quantum networking. This motivates our choice to consider implementing \dqnum in sequential quantum networks where for tractability we consider separable utility functions. We further assume a setting with no losses in classical control information exchange. Extending the framework to account for classical losses require careful investigation and is left for future investigation.

In a sequential quantum network, one node is designated as the source and another as the destination (or the sink), with entanglement distribution initiated at the source and progressing hop-by-hop towards the sink. Entanglement swaps are performed at each intermediate node along the path only after swaps have been completed at the preceding nodes. For example, consider a path consisting of intermediate nodes $v_0, v_1, v_2 \in V$ connecting the source $A_r$ with the sink $B_r$ (Figure \ref{fig:seq-qnum}a). A sequential quantum network distributes E2E entanglement to $(A_r, B_r)$ by performing an entanglement swap at a node $v_i$, only after  swaps have been completed at the previous node $v_{i-1}$.

Sequential quantum networks can use quantum datagrams (\emph{q-datagrams}) to facilitate E2E entanglement generation between two end-nodes $(A_r, B_r)$ \cite{bacciottini2024leveraging}. A q-datagram carries classical information such as destination end-node address ($B_r$) and previous Bell state measurement outcomes. As it moves through the network, each quantum switch uses the q-datagram to determine the next link on which to generate an LLE. It then performs entanglement swaps, thereby moving the entanglement towards the sink, while the source end node $A_r$ holds on to its half of the entangled state. The q-datagram is updated and forwarded hop-by-hop as swaps are performed, until it reaches the sink $B_r$. Upon successful delivery, $B_r$ performs necessary corrections and sends an acknowledgment (ACK) back to the source node $A_r$ to confirm E2E entanglement generation.

\begin{figure}[tb]
    \centering
    \includegraphics[width=\linewidth]{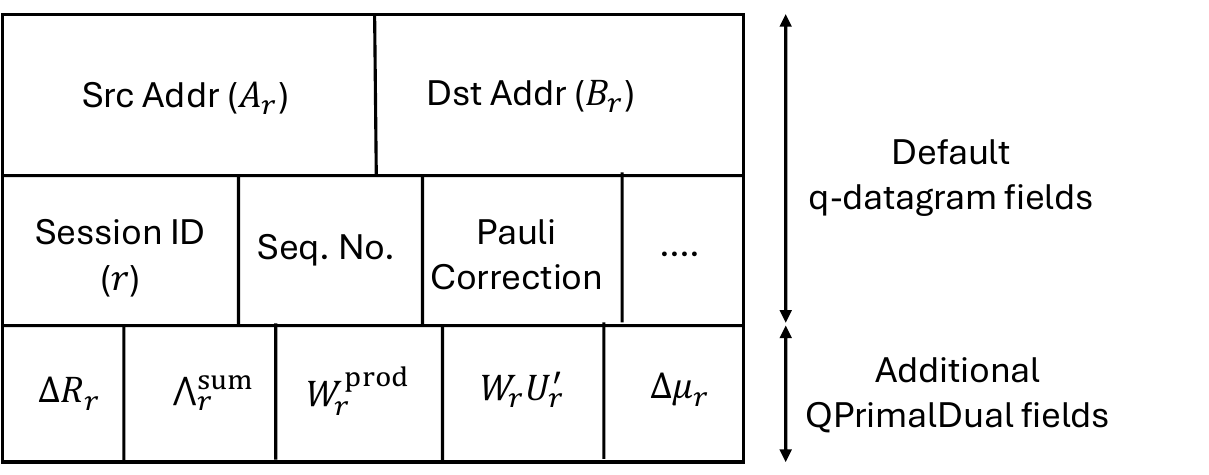}
    \caption{q-datagram header (classical) fields for \dqnum\;implementation in sequential quantum networks.}
    \label{fig:header}
    \vspace{-5mm}
\end{figure}

\subsection{q-datagrams and ACKs}\label{sec:qfields}

The deployment of \dqnum within sequential quantum networks is achieved by augmenting the q-datagram header \cite{bacciottini2024leveraging} with several additional fields as shown in Figure \ref{fig:header}. The original header includes fields like source and destination addresses, sequence number, session IDs, and Pauli correction bits. The newly added fields specific to \dqnum\;are described below.

\begin{enumerate}
    \item $\Delta R_r$: Change in session rate (q-datagram generation rate) since the last q-datagram was sent.
    \item $\Lambda_r^{\text{sum}}$: Cumulative sum of $\lambda_l$ of all traversed links as q-datagram travels to sink (updated hop-by-hop).
    \item $W_r^{\text{prod}}$: Cumulative product of $w_l$ of all traversed links as q-datagram travels to sink (updated hop-by-hop).
    \item $W_rU_r^{\prime}$: Stores $W_r\partial U_r(R_r,\vec{w}_r)/\partial W_r$ (See Section \ref{sub:controller-state} for the motivation behind including this field in the header), where $U_r^{\prime} = \partial U_r(R_r,\vec{w}_r)/\partial W_r$. Here, the end-to-end Werner parameter $W_r$ is obtained from the ACK.
    \item $\Delta\mu_r$: Change in $\mu_r$ since the last q-datagram was sent.
\end{enumerate}

Among the above fields, only $\Lambda_r^{\text{sum}}$ and $W_r^{\text{prod}}$ are updated hop-by-hop as the q-datagram traverses the network. The ACK carries the final aggregated values of these two fields back to the session source.
\subsection{Controller state}\label{sub:controller-state}
In a sequential quantum network (e.g., \cite{bacciottini2024leveraging}), link controllers are responsible for LLE generation and q-datagram processing. They are situated at the midpoint of a link, or at one of its endpoints. To support the \dqnum\;framework, we  extend their functionality to update link prices ($\lambda_l$) and Werner parameters ($w_l$) for each link $l \in \cL$. We also introduce a session controller\footnote{Referred to as a {\em source controller} in \cite{bacciottini2024leveraging}.} at each source node to update session rates ($R_r$) and E2E fidelity prices ($\mu_r$) for each $r \in \mathcal{R}$ (see Figure~\ref{fig:seq-qnum-a}). 

To perform \dqnum\;updates, both session and link controllers need to maintain certain local state variables. These variables allow the controllers to perform distributed updates without requiring full knowledge of the global network state.
\begin{itemize}[leftmargin=*]
\item Session controller state: at each source node, a session controller is responsible for tracking $R_r$, $\mu_r$ and $W_r.$
\item Link controller state: each link controller maintains its link specific primal ($w_l$) and dual ($\lambda_l$) variables, both of which are initialized to positive random values within their respective feasible ranges. Each link controller also maintains session specific $f_l(R_r,\vec{w}_r)$. Additionally, each link controller maintains three auxiliary variables: 
\begin{itemize}
\item $R_l^{\text{sum}}:$ initialized to zero, which stores $\sum_{r:l\in r}R_r$, required for updating $\lambda_l$ as per \eqref{dual-lambda},
\item $M_l^{\text{sum}}:$ initialized to zero, which stores$\sum_{r:l\in r}\mu_r$ (needed for updating $w_l$ per \eqref{dual-w}), and
\item $f_l^{\text{sum}}:$ initialized to zero, which holds$\sum_{r:l\in r}f_l(R_r,\vec{w}_r)$ (needed for updating $w_l$ per \eqref{dual-w}). Under the separability assumption, $f_l$ is only a function of $W_r$ and $w_l.$
\end{itemize}
\end{itemize}

Note that session controllers do not track per-link variables such as $\lambda_l$ and $w_l$. Rather, they obtain aggregate quantities involving these values from the ACK messages sent by the sink node. These ACKs carry cumulative metrics (e.g., product of $w_l$'s and sum of $\lambda_l$'s) that were gathered hop-by-hop by the q-datagram as it moved along the path (see Figure \ref{fig:seq-qnum}b). Link controllers only need to store the session-level $f_l(R_r,\vec{w}_r)$ values to compute $f_l^{\text{sum}}$. Also, note that 
    \begin{align}\label{eq:f_l_ratio}
    f_l(R_r,\vec{w}_r) &=\frac{\partial U_r(R_r,\vec{w}_r)}{\partial w_l} \nonumber \\
    &= \frac{\partial W_r}{\partial w_l}\cdot\frac{\partial U_r(R_r,\vec{w}_r)}{\partial W_r} = \frac{W_rU_r^{\prime}}{w_l}.
    \end{align}
Thus, link controllers do not require knowledge of the utility functions. They obtain $W_rU_r^{\prime}$ from the q-datagram header and compute $f_l(R_r,\vec{w}_r)$ by applying \eqref{eq:f_l_ratio}. The sums $M_l^{\text{sum}}$ and $R_l^{\text{sum}}$ can be incrementally updated based on the changes in per-session values conveyed through q-datagram headers. As q-datagrams traverse a link, they carry updated values of $\Delta R_r$ and $\Delta\mu_r$, enabling each link controller to update the cumulative sums. This design ensures that both types of controllers can operate based on locally maintained states, minimizing inter-controller communication.
\subsection{Roles of the session and link controllers, and the sink}
The roles of the session and link controllers and the sink are described below based on how they respond to different events. 

\noindent{\textbf{Session controller} (session $r$):}
\begin{enumerate}[leftmargin=*]
\item {\it On session initiation:} 
\begin{enumerate}
\item Initialize $R_r, \mu_r$ and $W_r$ to random feasible values. Set $\Delta R_r = R_r$ and $\Delta \mu_r = \mu_r$. Also, set $W_rU_r^{\prime} = W_r\partial U_r(R_r,\vec{w}_r)/\partial W_r$. Generate a q-datagram header that includes additional fields (see Section \ref{sec:qfields}) required by the \dqnum\;framework and forward it to the first link controller along the path. 
\item Start a timer of duration $1/R_r$, ensuring that the average rate of q-datagram generation aligns with the target session rate $R_r.$ We refer to this timer as the {\it generation timer} associated with the session.
\end{enumerate}
\item {\it On generation timer expiry:} Generate a q-datagram header as required by the \dqnum\;framework and transmit it to the first link controller along the path. Then, set a new generation timer of duration $1/R_r$.
\item {On receiving an ACK from the sink:}
\begin{enumerate}
\item Compute $R_r$ according to \eqref{dual-rate}, using $\Lambda_r^{\text{sum}}$ from the latest available ACK, then calculate $\Delta R_r$ by taking the difference between the new and previous session rates and forward it in the next q-datagram.
\item Update $W_r \leftarrow W_r^{\text{prod}}$ (from the ACK).
\item Compute $\mu_r$ according to \eqref{dual-mu} using $W_r^{\text{prod}}$, and then calculate $\Delta \mu_r$. For the bi-level update scheme, $\mu_r$ is updated only if the outer-level update is due.
\item Generate a q-datagram with the necessary fields and forward it to the next link controller and reset the generation timer to $1/R_r.$
\end{enumerate}
\item {\it On session termination:} Create a special classical control message and include the current value of $R_r$ and $\mu_r$. Then release all resources allocated to the session.
\end{enumerate}

\noindent{\textbf{Link controller} $l$:}
\begin{enumerate}[leftmargin=*]
\item {\it On receiving a q-datagram (say from session $r$):} Process it by generating an LLE and assigning it to the q-datagram. The expected generation time of this LLE is determined by the current value of the link Werner parameter $w_l$. If the LLE generation source is currently occupied with serving another q-datagram, the incoming q-datagram is placed in a queue. We assume a First-In-First-Out (FIFO) queuing discipline, with q-datagrams served in their arrival order. Once a q-datagram is selected for service, the link controller updates its internal state and the q-datagram header as follows:
\begin{enumerate}
\item Read $\Delta R_r$ from the q-datagram header. 
Update $R_l^{\text{sum}} \leftarrow R_l^{\text{sum}} + \Delta R_r.$ Use $R_l^{\text{sum}}$ to update $\lambda_l$ via \eqref{dual-lambda}.
\item Read $W_rU_r^{\prime}$ and $\Delta\mu_r$ from the q-datagram. Update $M_l^{\text{sum}} \leftarrow M_l^{\text{sum}} + \Delta\mu_r.$ Use $W_rU_r^{\prime}$ and other stored values of $f_l(R_{r^{\prime}},\vec{w}_{r^{\prime}}), r^{\prime} \in \cR\setminus \{r\}$ to compute $f_l^{\text{sum}}$. Use $f_l^{\text{sum}}$ and $M_l^{\text{sum}}$ to update $w_l$ in \eqref{dual-w}. Update $f_l(R_{r},\vec{w}_{r})$ locally. As per bi-level update scheme, update $w_l$ only if the outer-level update is due.
\item Compute $W_r^{\text{prod}} \leftarrow W_r^{\text{prod}} \cdot w_l$ and update the q-datagram field accordingly.
\item Compute $\Lambda_r^{\text{sum}} \leftarrow \Lambda_r^{\text{sum}} + \lambda_l$ and update the q-datagram.
\item Forward the q-datagram header to the next link controller or the sink along the path.
\end{enumerate}
\item {\it On receiving session termination request:} Retrieve the values of $R_r$ and $\mu_r$ and update the running sums by subtracting the session's contribution: $R_l^{\text{sum}} \leftarrow R_l^{\text{sum}} - R_r$ and $M_l^{\text{sum}} \leftarrow M_l^{\text{sum}} - \mu_r$.
\end{enumerate}

\noindent {\bf The sink:} {\it On receiving a q-datagram (say from session $r$):} Read $\Lambda_r^{\text{sum}}$ and $W_r^{\text{prod}}$ fields from the header and send them along with an ACK to the session controller of session $r$.

\subsection{\dqnum\;extensions}\label{subsec:ext}
The \dqnum\; implementation presented in the previous section follows the theoretical formulation presented in Section~\ref{sec:dqnum} and, under ideal conditions, inherits its stability guarantees. However, a practical deployment raises additional concerns regarding scalability and robustness to losses and quantum noise. In this section, we discuss these concerns and propose three extensions to address them.

\subsubsection{\dqnumapproax}
In this variant, we approximate the sum of session rates at each link controller. This algorithm variant is called \dqnumapproax, and it eliminates the $\Delta R_r$ field from the q-datagram header. Every link controller estimates $R_l^{\text{sum}} =  \sum_{r : l \in r} R_r$ by maintaining an exponential average $T_{\text{int}}$ of the q-datagram interarrival times. Let $t'$ and $t''$ be the arrival times of the two most recent q-datagrams at link controller $l$. Then, we track the sum of session rates using the following update rules, applied at each link controller $l$ whenever a new q-datagram arrives:
\begin{align}
    T_{\text{int}} &\leftarrow \alpha T_{\text{int}} + (1-\alpha)(t'' - t'), \;\; R_l^{\text{sum}} \leftarrow 1/T_{\text{int}},\label{eq:approx-O}
\end{align}
where $\alpha \in (0,1)$ is a tunable parameter set to $0.9$ in our simulations. 
The approximation is valid as long as q-datagram interarrival times are relatively stable. It is also possible to use more sophisticated techniques to estimate the sum of session rates, but we leave this for future work. The \dqnumapproax\;variant is inherently more robust to network dynamics and malicious users, as it does not need to trust the session rate updates provided by the users.

\subsubsection{\dqnumpatch} 
Another concern is the robustness of \dqnum\;to decoherence. In sequential quantum networks, the LLEs are stored in quantum memories at a node until entanglement swapping is performed. The coherence time of these memories limits the time available for generating E2E entanglement, as fidelity degrades over time. To address this issue, we propose an enhancement to \dqnum\;that accounts for memory decoherence by monitoring queueing delays. We refer to this extension as decoherence-aware \dqnum\;(\dqnumpatch).

We assume that all quantum memories in the network have the same coherence time $T_c$, with decoherence modeled as time-dependent depolarizing noise with probability $1 -e^{-t/T_c}$. Our approach is based on modeling each link controller as an $M/M/1$ queue  \cite{deandrade2024analysisquantumrepeaterchains}, such that the expected waiting time of a q-datagram at each link is $T_W^l= (d_l(1-w_l) - \sum_{r:l\in r}R_r)^{-1}$. The $M/M/1$ assumption has already been identified as a useful tool for analyzing the performance of sequential quantum networks in the literature \cite{deandrade2024analysisquantumrepeaterchains}.
We modify the original QNUM formulation by replacing the link capacity constraint \eqref{eq:optimal_qnum_cap} with 
\begin{equation}
    \sum_{r:l\in r}R_r  \leq  d_l(1-w_l) - \frac{G}{T_c}, \quad \forall{l \in \cL},\label{eq:optimal_qnum_cap_patch}
\end{equation}
where $G>1$ is a tunable parameter. $T_W^l \leq T_c/G$ is now enforced by \eqref{eq:optimal_qnum_cap_patch}. \dqnumpatch\;introduces a rate penalty of $G/T_c$ which can be neglected if $d_l(1 -w_l) \gg G/T_c$ or can still be tolerated unless the link's capacity $d_l(1 -w_l) \approx 1/T_c$. If the last condition is met, then the utility of the quantum network is limited as the time it takes to generate a new LLE would be comparable to the LLE's coherence time. Moreover, queueing can occur in any quantum network architecture unless all devices and users are perfectly synchronized. This makes our proposed extension valuable beyond sequential quantum networks. In our simulations, we set $G=50$.

We also define the extension \emph{\dqnumpatchapprox}\, obtained by applying the method proposed in \dqnumapproax\, to \dqnumpatch\, instead of \dqnum.
\subsubsection{\dqnumpi}
An alternative solution to tackle decoherence is to adopt a two-step approach. Firstly, we use \dqnum\;until the control variables $\vec{R}$, $\vec{w}$, $\vec{\lambda}$ and $\vec{\mu}$ converge to their steady state. Then, we fix $w_l$ on all links and switch to a quantum network protocol designed to mitigate decoherence. 
As an example, (after convergence) consider replacing the session controllers of \dqnum\;with Quantum Transmission Control Protocol (QTCP) congestion controllers from \cite{bacciottini2024leveraging}, and substituting the link controllers with Proportional-Integral (PI) controllers \cite{bacciottini2024leveraging}. QTCP saturates physical link capacity by design, and can thus be used without performance loss when Werner parameters $w_l$ are frozen to their optimal values. On the other hand, PI controllers seamlessly integrate with QTCP to limit sending rates and prevent congestion in the quantum network. This engineering solution realizes a similar trade-off between rate loss and decoherence mitigation as \dqnumpatch, while also providing the desirable plug-and-play feature of QTCP, which allows for easy integration with other quantum network protocols. With this solution, it becomes necessary to detect changes/failures in the quantum network and re-run \dqnum\;to update the $w_l$ variables. We refer to this approach as \dqnumpi.

\subsection{Handling q-datagram losses}
Even with reliable classical communication and deterministic swaps, a q-datagram might still be lost if nodes have limited quantum memory. Both \dqnum\;and its extensions need to account for the possibility of q-datagram losses. For example, if a q-datagram is dropped just before reaching the link controller $l$, the link states (e.g., $R_l^{\text{sum}}$ and $M_l^{\text{sum}}$) of upstream links (i.e., those preceding $l$) would have already been updated, while  downstream link states (including $l$ and beyond) would remain unchanged. This leads to state inconsistencies which we address in Appendix~\ref{sub:qdatagram-loss-appendix}.

\section{Performance Evaluation}\label{sec:perf-eval}
\begin{figure}[t]
\centering
\begin{minipage}[t]{0.35\linewidth}
\centering
\includegraphics[width=\linewidth]{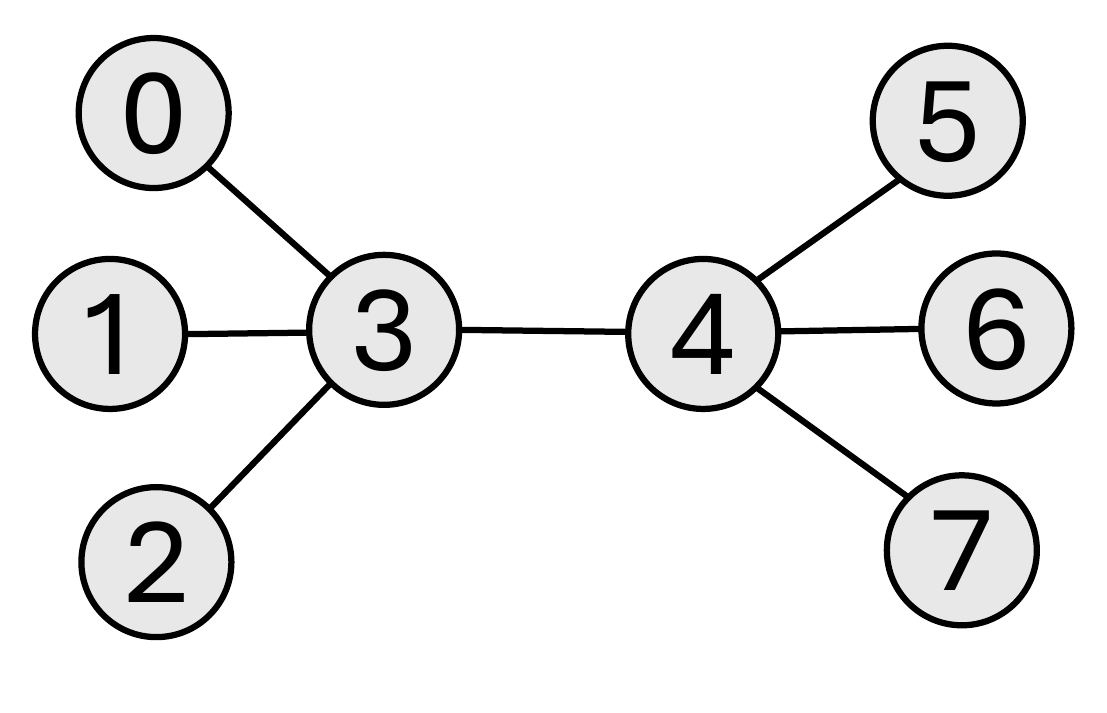}
\subcaption{Dumbbell}\label{fig:dumbbell}
\end{minipage}
\hspace{0.2cm}
\begin{minipage}[t]{0.6\linewidth}
\centering
\includegraphics[width=\textwidth]{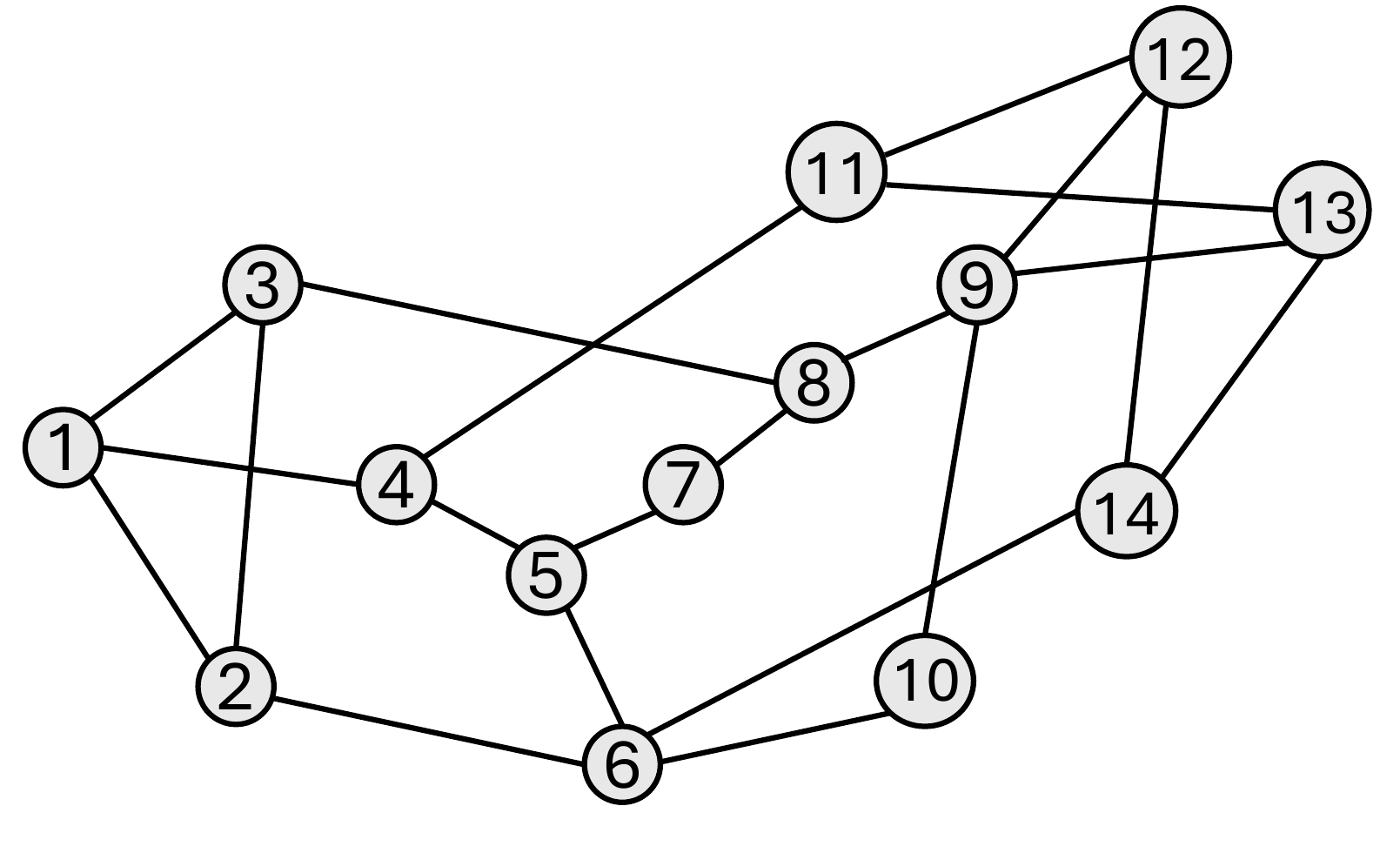}
\subcaption{\textit{NSFNet}}\label{fig:nsfnet}
\end{minipage}
\caption{The two simulated topologies.}
\label{fig:topologies}
\end{figure}

The goal of this section is fourfold: (i) to demonstrate that \dqnum\;remains stable under realistic conditions; (ii) to confirm that \dqnum\; converges to the optimum; 
(iii) to evaluate the scalability and robustness of \dqnum\;to dynamic scenarios, including link failures and dynamic workloads; (iv) to evaluate the effectiveness of the \dqnum\;extensions against quantum memory decoherence.

\subsection{Simulation Setup}\label{subsec:sim-setup}

\begin{table}[tb]
\centering
\begin{tabular}{ll}
\toprule
\multicolumn{2}{c}{\textbf{System-wide assumptions}} \\
\midrule
\textit{Quantum operations} & Deterministic, noiseless swaps\\
\textit{Decoherence model} & Exponential depolarizing decay\\
\textit{LLE generation} & i.i.d. geometric inter-arrival times \\
\textit{Classical communications} & Reliable, speed-of-light propagation delay \\
\textit{Quantum/Classical routing} & Shortest path \\
\end{tabular}

\vspace{1em}
\scriptsize
\begin{tabular}{lll}
\toprule
\textbf{Parameter} & \textbf{Value} & \textbf{Description} \\
\midrule
\multicolumn{3}{l}{\textit{Hardware parameters}} \\
\midrule
$\chi_l$ & $100\,\mathrm{kHz}$ & Photon source repetition rate \\
$N_{\text{mem}}$ & $50$ & Memory qubits per node per attached link \\
$T_c$ & $1\,\mathrm{s}$ & Memory coherence time (if considered) \\
$\eta_l$ & $0.25 \cdot e^{-(L_l/2)/22}$ & Accounts for fiber loss, freq. conversions, ...\\
\midrule
\multicolumn{3}{l}{\textit{\dqnum \,parameters}} \\
\midrule
$T_{\text{outer}}$ & $10$ & Outer-loop period (\# iterations) \\
$k_{\lambda_l}$ & $[10^{-6}, 10^{-5}]$ & Step size for link dual variables $\lambda_l$ \\
$k_{\mu_r}$ & $10^{-2}$ & Step size for session dual variables $\mu_r$ \\
$k_{w_l}$ & $[10^{-5}, 10^{-4}]$ & Step size for Werner parameters $w_l$\\
$w_l(t=0)$ & $0.967$ & Initial Werner parameter per link \\
$w_l(\text{QTCP})$ & $0.967$ & Fixed Werner parameter in baseline protocol \\
\bottomrule
\end{tabular}
\caption{Simulation configuration.}
\label{tab:sim-params}
\vspace{-5mm}
\end{table}
We consider two network topologies: (i) an $8$ node, $7$ link dumbbell topology (Figure \ref{fig:dumbbell}). It is one of the widely used topologies in (classical) networking research \cite{winstein2013tcp, floyd2003internet, wang2013improved} for testing performance under congestion and bottlenecks; and (ii) the \textit{NSFNet} topology \cite{Mills_1987} (Figure \ref{fig:nsfnet}), which consists of $14$ nodes and $21$ links. All links in the dumbbell topology are of equal length $80\,\mathrm{km}$ unless otherwise specified, while the \textit{NSFNet} links have different lengths as specified in \cite{Mills_1987}, downscaled by a factor of $25$ so that the photon loss probability is tolerable. 

The default configuration for the simulations is summarized in Table \ref{tab:sim-params}. The coherence time value $T_c=1\ \text{s}$ is optimistic, but supported by recent results \cite{Ma2021, Park2022}. Other configurations that we evaluated incorporate attempt rates $\chi_l$ in the order of $1$ GHz \cite{Chen_2023} and shorter coherence times, leading to qualitatively similar results to those presented in this section. We consider the utility functions $U^{\text{SKR}}_r$ and $U^{\text{neg}}_r$ defined in \eqref{eq:skr}, \eqref{eq:neg}, and label them SKR and NEG for readability. For an easier interpretation of the results, we plot the absolute values of SKR and NEG and not their logarithms.

Throughout this section, we refer to the \textit{convergence time} of \dqnum. In simulations, this is the earliest time after which the $10$ s moving average of the aggregate utility remains within $5$\% of its steady-state value.

For steady-state performance, we plot the average over $32$ independent Monte Carlo simulations. Each simulation executes for roughly ten times the \dqnum\; convergence time, corresponding to about $160$ seconds in most cases.
In the dumbbell topology experiments, the workload consists of six user sessions established between node pairs \((i, i+3)\) for \(i \in \{0,1,2\}\) (see Figure~\ref{fig:dumbbell}). For each pair, two sessions are defined: one with node \(i\) acting as the session controller and the other with node \(i+3\) as the session controller. We exploit the \textit{NSFNet} topology to test \dqnum\;with a variable number of sessions between random pairs of nodes.

The QTCP protocol \cite{bacciottini2024leveraging} is selected as the baseline to compare against the \dqnum\;variants. The Werner parameters for QTCP are fixed at $w_l\approx0.967$, $l\in L$ obtained by maximizing the secret key rate for the dumbbell topology locally and independently at each link controller. In other words, this is the optimal Werner parameter under the constraint of having a homogeneous, fixed configuration.

\subsection{Robustness of the bi-level update scheme}\label{subsec:bilevel-robustness}
\begin{figure}[tb]
    \centering
    \includegraphics[width=\linewidth]{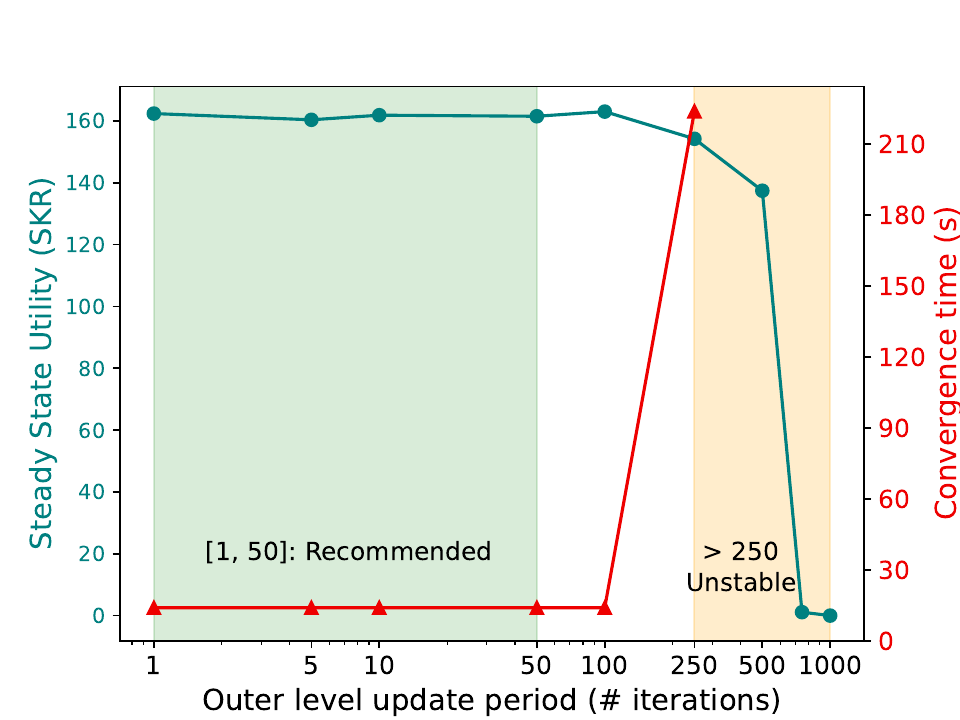}
    \caption{Convergence of the hierarchical bi-level update scheme on the dumbbell topology with different values of the outer-level update period. Red markers (convergence time) are not shown if the algorithm does not converge.}
    \label{fig:bi-level}
\end{figure}

We evaluate the convergence properties of the bi-level approach in Figure \ref{fig:bi-level}. Specifically, we examine the impact of varying the outer-level update period $T_{\text{outer}}$ on the steady-state utility and the convergence time. While the theoretical convergence guarantees provided in Section~\ref{subsec:stability} hold for outer-level update periods equal to one iteration, our simulation results demonstrate that the algorithm continues to exhibit stable convergence for larger outer level update periods. We evaluated convergence properties under different link lengths on the dumbbell topology and on \textit{NSFNet}. We show in Figure \ref{fig:bi-level} some representative results on the dumbbell topology. We observe that an outer level period $T_{\text{outer}} \in [1, 50]$ ensures convergence, while $T_{\text{outer}} \geq 250$ might lead to instability.

\subsection{Steady-State Performance}
\begin{figure*}[tb]
\centering
\begin{minipage}{0.45\textwidth}
\includegraphics[width=1\textwidth]{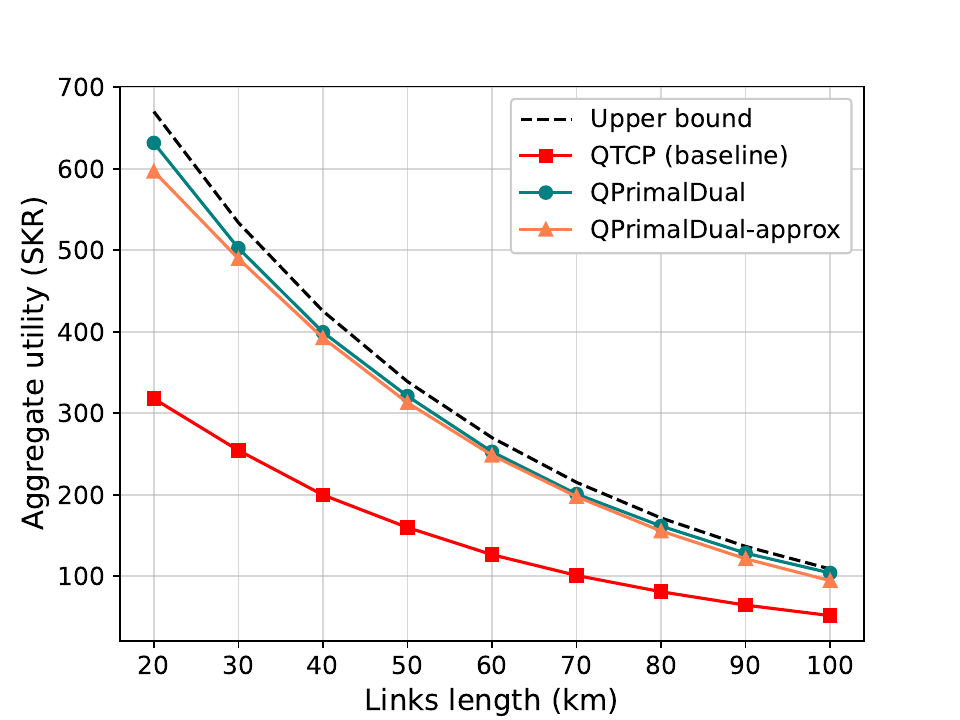}
\subcaption{No decoherence}\label{fig:utility_dumbbell_a}
\end{minipage}
\begin{minipage}{0.45\textwidth}
\includegraphics[width=1\textwidth]{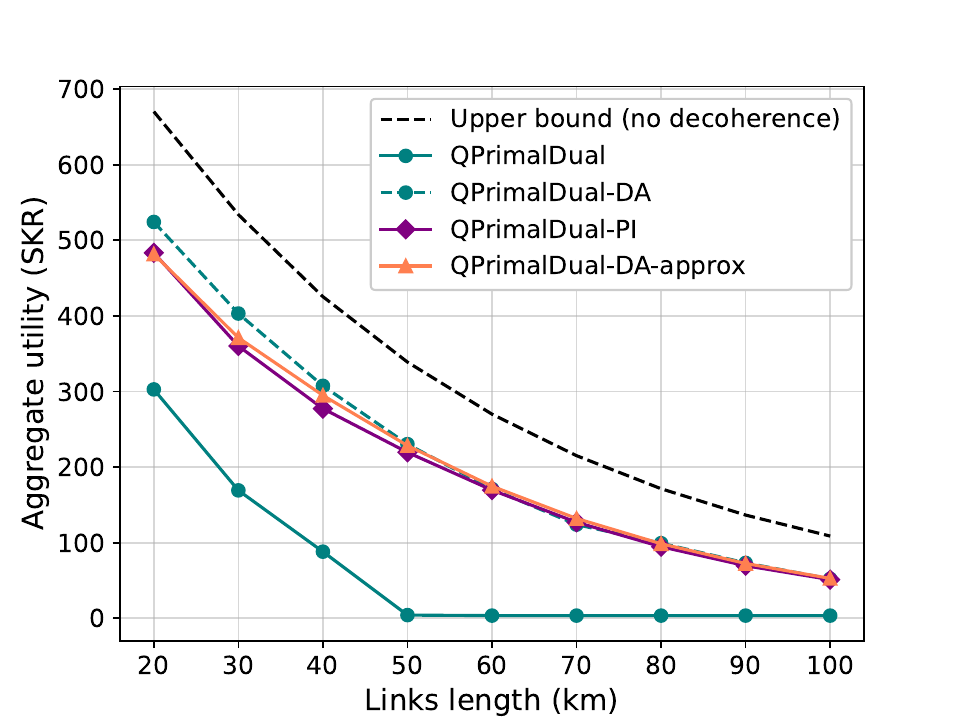}
\subcaption{Coherence time $T_c=1$ s}\label{fig:utility_dumbbell_b}
\end{minipage}
\caption{Steady-state performance of the baseline and \dqnum\;variants on the dumbbell topology as a function of link length. The upper bound is obtained by analytically solving QNUM with the same parameters. In case of decoherence, the upper bound, which is the same for both (a) and (b), is inevitably loose as it is computed assuming that qubits are immediately measured without spending time in memory. Confidence intervals are smaller than or comparable to marker sizes.}
\label{fig:utility_dumbbell}
\end{figure*}

In Figure \ref{fig:utility_dumbbell}, we evaluate the steady-state performance of \dqnum\;on the dumbbell topology, with and without decoherence. With perfect memories (Figure \ref{fig:utility_dumbbell_a}), the performance of \dqnum\;and \dqnumapproax\;(teal and orange lines) follow the analytical upper bound (dashed line) with less than a $5$\% gap in aggregate secret key rate. The small gap is due to sporadic q-datagram losses caused by congestion (see Appendix \ref{sub:qdatagram-loss-appendix} for details). In comparison, QTCP consistently underperforms, showcasing the advantage of a protocol that globally and jointly optimizes session rates and link Werner parameters. 

When we introduce decoherence (Figure \ref{fig:utility_dumbbell_b}), the performance of \dqnum\;(solid teal) drops significantly. Link capacities saturate and q-datagrams experience long waiting times and, consequently, decohere. However, both \dqnumpatch\;and \dqnumpi\;(dashed teal and purple) are able to restore performance. While neither approach fully closes the gap to an upper bound (independent of decoherence), both provide practical solutions that maintain high aggregate utility with predictable rate loss. We also plot the SKR obtained by \dqnumpatchapprox\;(orange) that combines the two extensions. It achieves similar performance to \dqnumpatch\;while reducing communication overhead.

\subsection{Fast Reaction to Failures}

\begin{figure}[tb]
\centering
\includegraphics[width=\linewidth]{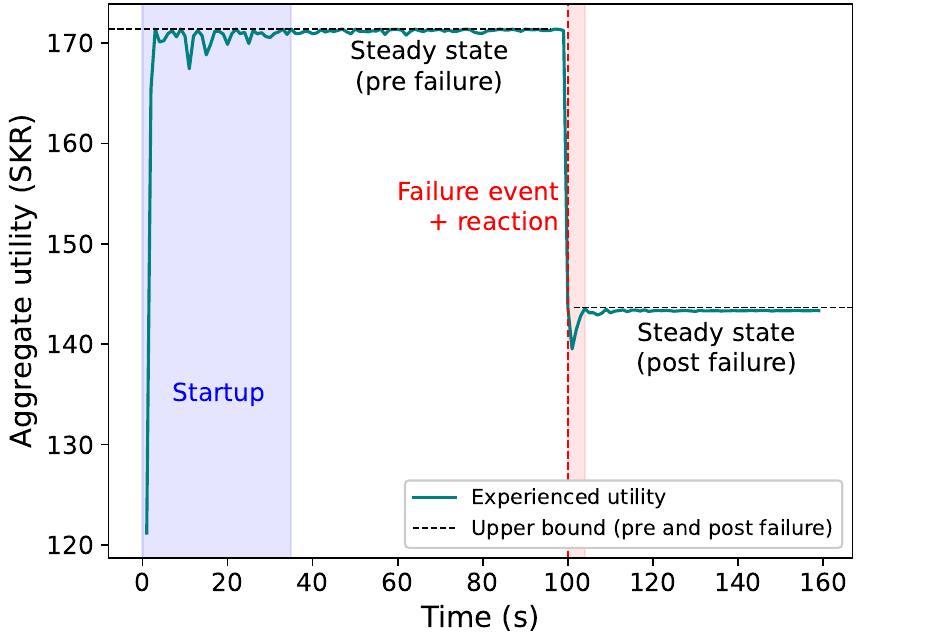}
\caption{Simulation trajectory of the network aggregate utility using \dqnum\;on the dumbbell topology. The protocol converges to the optimal secret key rate and quickly recovers to a new optimum after a failure of link $1$-$3.$}
\label{fig:failure-dumbbell}
\end{figure}

We consider a representative scenario on the dumbbell topology where we bootstrap \dqnum\; and, after some time, a physical link fails. We plot in Figure \ref{fig:failure-dumbbell} how the aggregate SKR evolves over time before and in reaction to the failure. As shown in Figure \ref{fig:failure-dumbbell}, \dqnum\;goes through a startup phase whose duration depends on the step sizes and the initial conditions. During this phase, \dqnum\; exhibits transitory oscillations, and then converges to the optimal value (dashed black line). To evaluate \dqnum’s ability to adapt to dynamic network conditions, we simulate a failure by artificially disabling the $1$--$3$ link at time $t = 100\,\mathrm{s}$. We observe that the protocol is able to quickly adapt and converge to the new optimum. The same behavior occurs if, instead of a link failure, the two sessions traversing the failed link were to suddenly leave the network, as the link controller would not receive any more q-datagrams from those sessions.
\begin{figure*}[tb]
\centering
\begin{minipage}{0.45\textwidth}
\includegraphics[width=1\textwidth]{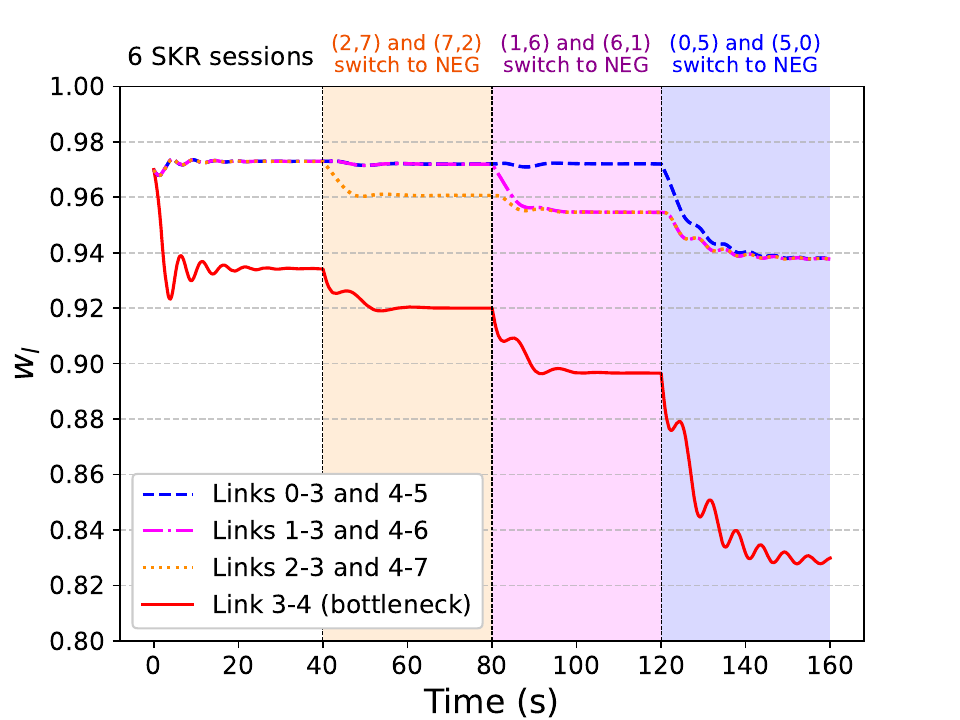}
\subcaption{\dqnum}\label{fig:wl_dumbell-a}
\end{minipage}
\begin{minipage}{0.45\textwidth}
\includegraphics[width=1\textwidth]{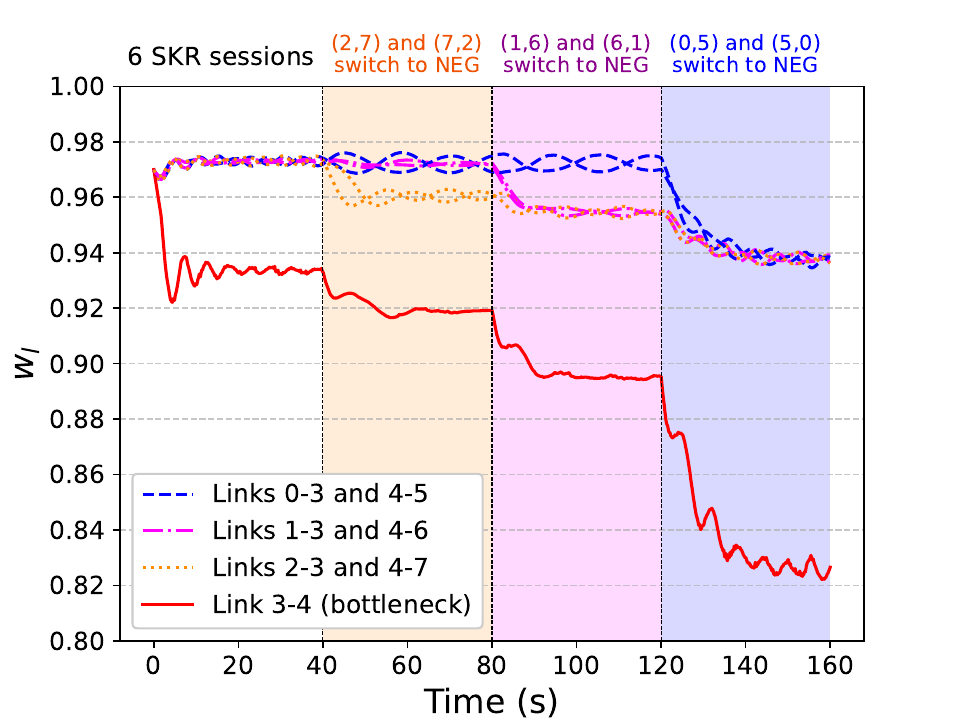}
\subcaption{\dqnumapproax}\label{fig:wl_dumbell-b}
\end{minipage}
\caption{A comparison of (a) \dqnum\;and (b) \dqnumapproax\;under dynamic workloads. The plots show the trajectories of the link-level Werner parameters $w_l$ over a representative Monte Carlo simulation. During the first $40$ s (white areas), all six sessions use SKR utility. Then, every $40$ s two sessions switch to NEG, until all sessions use the same utility (blue areas). The background colors highlight which sessions have last switched to a different utility.}
\label{fig:wl_dumbell}
\end{figure*}

\subsection{Dynamic workloads}
Figure \ref{fig:wl_dumbell} illustrates how \dqnum\;reacts to a dynamic workload. Background colors in the plot indicate different workload phases, with user sessions switching between SKR and NEG. As more users switch to NEG, we observe that the link-level Werner parameters $w_l$ quickly decrease to lower values, which is consistent with the fact that secret key rate is more sensitive to fidelity than entanglement negativity \cite{Vardoyan23QNUM}. We also observe that the bottleneck link $3$--$4$ (red solid line) maintains a lower Werner parameter (and thus a higher capacity) to accommodate all six user sessions. We observed similar results with other dynamic workloads where the number of user sessions changes over time.

Figure \ref{fig:wl_dumbell-b} also provides insight into why \dqnumapproax\;performs slightly worse than \dqnum. The approximations lead to a less accurate and slightly fluctuating estimate of the session rates, which cause oscillations of the link-level Werner parameters $w_l$ around their optimal values. However, as confirmed by all the simulated scenarios, the performance gap is small and the approximation simplifies the implementation, which makes \dqnumapproax\;a viable alternative for large networks.

\subsection{Experiments on \textit{NSFNet}}
\begin{figure}[tb]
    \vspace{-5mm}
    \centering
    \includegraphics[width=\linewidth]{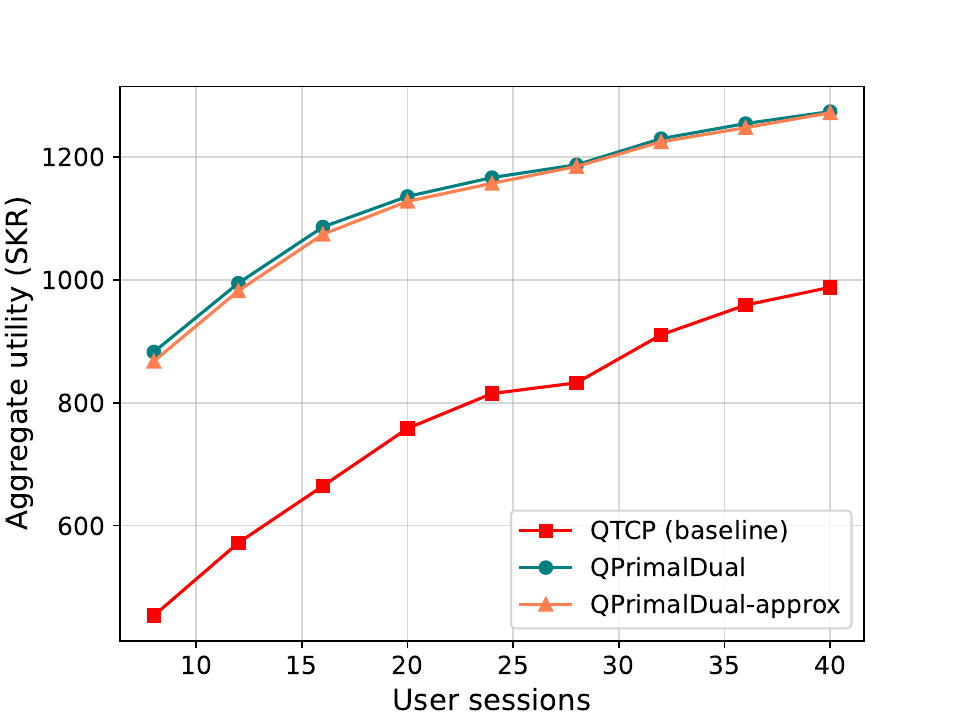}
    \caption{Expected steady-state performance of \dqnum\;on the \textit{NSFNet} topology as a function of the number of sessions.
    Each point is the average of $32$ sets of Monte Carlo simulations with randomly selected user sessions.}
    \label{fig:utility_NSFNet}
\end{figure}

In Figure \ref{fig:utility_NSFNet}, we show the expected steady-state performance of \dqnum\;and \dqnumapproax\;on the \textit{NSFNet} topology, as the number of user sessions increases. This set of simulations evaluates \dqnum's scalability to larger networks and its robustness to higher loads. Finding the QNUM optimum, either analytically or through a centralized solver, is difficult due to the non-concave nature of secret key rate utility and the complexity of a larger topology like \textit{NSFNet}. Hence there is no guarantee that \dqnum\;and \dqnumapproax\; converge to the global optimum. However, Figure \ref{fig:utility_NSFNet} confirms that the findings are consistent with those  obtained for the simpler dumbbell topology: \dqnum\;and \dqnumapproax\;achieve similar performance and both outperform QTCP in all scenarios. These results strongly suggest that the \dqnum\;algorithm remains efficient and robust at larger scales.

\section{Related Work}\label{sec:rel}
Most current resource allocation strategies in quantum networks adopt a centralized model \cite{Vardoyan23QNUM, beauchamp2025modular, gu2023esdi, skrzypczyk2021architecture, cicconetti2021request, Chakraborty_2020}. In centralized designs, a global controller is responsible for deciding the rate and fidelity at which LLEs should be generated and the rate at which E2E requests should be served. However, centralized solutions introduce a single point of failure and require complete knowledge of the network. Existing work on distributed resource allocation algorithms in quantum networks is limited and considered in \cite{zhao2024asynchronous, bacciottini2024leveraging}. These works adapt the classical TCP congestion control algorithm \cite{jacobson1988congestion} of the  Internet to quantum networks, managing E2E entanglement distribution similar to how TCP handles data packets in the Internet. However, these heuristic  approaches have two key limitations: (i) they do not account for the fidelity of the delivered E2E entangled states, (ii) they do not provide analytical guarantees, particularly with respect to convergence and stability. 

There has been an extensive amount of work on developing distributed solutions to the NUM problem for classical networks \cite{kelly1998rate, low1999optimization, chiang2007layering, chiang2006decomposition}. However, those approaches do not directly apply to quantum networks, where quantum effects like fidelity loss from entanglement swapping \cite{victora2023entanglement} and memory decoherence influence the end user utility and have no classical counterparts. Additionally, key network resources, such as LLEs, often face an inherent trade-off between their generation rate and fidelity depending on the underlying entanglement generation mechanism. For example, in protocols like the single click scheme \cite{humphreys2018deterministic, rozpedek2019building} or SPDC based schemes \cite{kok2000postselected, krovi2016practical}, tuning parameters to increase fidelity can decrease generation rate. Similar trade-offs exist in approaches involving entanglement distillation or quantum error correction \cite{dur2007entanglement}, where an increase in fidelity comes at the cost of reduced generation rate. Since both rate and fidelity influence the objective and constraint space, these additional considerations, in some cases, alter the convexity of the QNUM problem and require more coordination and information exchange for distributed implementations.
\section{Conclusion}\label{sec:dis}
We introduced a distributed framework for resource allocation in quantum networks, building on the foundational principles of  QNUM. Our proposed framework leverages two interacting controllers, located at end nodes and links, which operate using only local information and limited classical communication. We provided theoretical guarantees for stability under different convexity assumptions of the utility functions. Specifically, for separable and concave utility functions, we proved global asymptotic stability, while for non-concave cases, we established local asymptotic stability under separability and additional constraints. We further provided a practical protocol for implementing our algorithm on sequential quantum networks and proposed several protocol variants to deal with quantum memory decoherence and control overhead. Our simulations incorporated realistic feedback delays and empirically demonstrated convergence behavior across various settings. The stability analysis of our algorithm in the presence of feedback delays, as well as protocol level implementations for other entanglement swapping architectures remain our future work.

\bibliographystyle{IEEEtran}
\bibliography{bib}
\section{Appendix}
\subsection{Verification of Slater's Condition for QNUM}\label{sub:slater-appendix}
Consider the QNUM optimization problem defined in \eqref{eq:optimal_qnum_obj}-\eqref{nonnegativity-R}. Slater’s condition requires the existence of a strictly feasible point, i.e., $(\vec{R},\vec{w})$ such that
\begin{align*}
    \sum_{r:l \in r} R_r < d_l(1-w_l), \forall l \in \cL,
    \sum_{l \in r} \log w_l > K_r,\forall r \in \cR.
\end{align*}
Pick a small $\delta \in (0,1)$ and set $w_l = 1-\delta$ for all $l \in \cL$ such that $K^{\text{max}} < |r^{\text{min}}|\log(1-\delta)$. Here, $K^{\text{max}} = \max_r K_r,$ 
$r^{\text{min}} = \argmin_{r} |r|$ and $|r|$ is the number of links on route $r.$ Then for each session $r,$ pick a small positive $\epsilon$ and set $R_r = \epsilon\;\forall r \in \cR,$ such that $\epsilon < \min_l \;d_l\delta/N_l,$ where $N_l$ denotes the number of sessions that use link $l.$ $\delta$ and $\epsilon$ together ensure strict minimum E2E fidelity and capacity constraints. 

\subsection{Stability Analysis of \dqnum}
Consider the QNUM optimization problem defined in \eqref{eq:optimal_qnum_obj}-\eqref{nonnegativity-R}. Let $U = \sum_{r\in \cR}  U_r(R_r,\vec{w}_r).$ Given the assumption about separability, i.e., $\partial U_r/\partial R_r = f_r(R_r)$ is only a function of $R_r$ and that $\partial U_r/\partial w_l = f_l(\vec{w}_r)$ is only a function of $\vec{w}_r$ and the Lagrangian $\mathcal{A} (\vec{R}, \vec{w}, \vec{\lambda}, \vec{\mu}) $ defined in \eqref{eq:lag}, 
the following  differential equations enforce the KKT conditions: 
\begin{eqnarray}
\dot{w_l} &=&  \partial U/\partial w_l -d_l \lambda_l + \sum_{r:l \in r} \mu_r/w_l,  \;\; \forall l \in L \nonumber \\
\dot{\lambda_l} &=& \left[\sum_{r:l \in r} R_r - d_l(1-w_l)\right]_{\lambda_l}^{+},  \;\; \forall l \in L \nonumber \\
\dot{\mu_r} &=& \left[ K_r- \sum_{l:l \in r}\log(w_l)\right]_{\mu_r}^{+},  \;\; \forall r \in {\cal R}.
\label{eq: DE}
\end{eqnarray}
\noindent Here, for any $x$, $\dot{x}$ denotes the time derivative of $x$, and is approximated by $\dot{x} = x(t+1) - x(t)$ (see Section \ref{sub:dqnumPD}). The notation $[z]^+_a$ denotes the projection of $z$, such that the dual variable $a$ (e.g., $\lambda_l$ or $\mu_r$) remains non-negative.

\noindent
{\bf Concave Utility -- global asymptotic stability}: Let $U'_x$ denote $\partial U/\partial x$ and $U''_x$ denote $\partial^2 U/\partial x^2$. Without loss of generality, assume $d_l \equiv 1$ and consider  the candidate Lyapunov function
$$
V= \sum_{l\in L} V_{w_l}  + \sum_{r\in {\cal R}} V_{\mu_r} + \sum_{l\in L} V_{\lambda_l},
$$
where $V_{w_l}= \frac{1}{2} (w_l-w_{l}^*)^2$, $V_{\mu_r}= \frac{1}{2} (\mu_r-\mu_r^*)^2$ and $V_{\lambda_l}= \frac{1}{2} (\lambda_l-\lambda^*_l)^2$. Given $l \in L$ and $r \in {\cal R}$, we first compute the time derivative of $V_{w_l} + V_{\mu_r}$ along trajectories of (\ref{eq: DE}):
\begin{eqnarray*}
\dot{V}_{w_l} 
&=& (w_l-w_l^*)\dot{w_l}  \nonumber \\
&=& (w_l-w_l^*)(U'_{w_l}(\vec{w})-\lambda_l + \sum_{r:l \in r} \mu_r/w_l)  \nonumber \\
        &=& (w_l-w_l^*)(U'_{w_l}(\vec{w}) - U'_{w_l}(\vec{w}^*) +\sum_{r:l \in r} \mu_r/w_l  \nonumber \\
        &  & \mbox{} - \sum_{r:l \in r} \mu^*_r/w^*_l) - (w_l-w_l^*)(\lambda_l-\lambda_l^*);\nonumber \\
\dot{V}_{\mu_r} &=&  (\mu_r-\mu_r^*)\dot{\mu_r}= (\mu_r-\mu_r^*)(K_r-\sum_{l:l \in r}\log(w_l)) \nonumber \\
        &=& (\mu_r-\mu_r^*)\sum_{l:l \in r}\log(w^*_l/w_l) .
\end{eqnarray*}
Now, considering all the links and sessions:
\begin{align}
&\sum_{l\in L} \dot{V}_{w_l} + \sum_{r\in {\cal R}} \dot{V}_{\mu_r} \nonumber\\
= &\sum_{l\in L}(w_l-w_l^*)(U'_{w_l}(\vec{w}) - U'_{w_l}(\vec{w}^*)) \nonumber \\
& \mbox{} + \sum_{l\in L} \sum_{r: l \in r} \left[ \mu_r((w_l - w_l^*)/w_l + \log(w_l^*/w_l)) \right.\nonumber\\
& \mbox{} \left. +  \mu_r^*((w_l^* - w_l)/w_l^* + \log(w_l/w_l^*)) \right] \nonumber \\
& \mbox{} - (\vec{w}-\vec{w}^*)^T(\vec{\lambda} - \vec{\lambda}^*).\label{eq:two_dotv_sums}
\end{align}

Because of the first order condition of concavity (Chapter 3, \cite{boyd2004convex}), the following two inequalities hold for any $\vec{w}$ and $\vec{w}^*$: 
\begin{align}
D(\vec{w}^*) &\le D(\vec{w}) + \Delta D(\vec{w})^{T}(\vec{w}^* - \vec{w}),\nonumber\\
D(\vec{w}) &\le D(\vec{w}^*) + \Delta D(\vec{w}^*)^{T}(\vec{w} - \vec{w}^*),\nonumber
\end{align}
where $\Delta D(\vec{w}) = [U'_{w_1}(\vec{w}), U'_{w_2}(\vec{w}),...,U'_{w_l}(\vec{w}),...]^T$ and $D(\vec{w}) = \sum_{l\in L}\int U'_{w_l}(\vec{w}) \,dw_l .$ Adding the above two inequalities, we get:
$(\Delta D(\vec{w}) -  \Delta D(\vec{w}^*))^T(\vec{w}-\vec{w}^*) \le 0 $. Therefore, the first term on the right-hand side of \eqref{eq:two_dotv_sums} is non-positive.
Each of the summands of the double sum  in \eqref{eq:two_dotv_sums} are of the form:
$$
g(x_i, x_j) = (x_i-x_j)/x_i + \log(x_j/x_i),
$$
which is less than or equal to zero for all $0 < x_i, x_j < 1$.

We now consider the time derivative of $V_{\lambda_l}$. First, let $y_l :=\sum_{r:l \in r} R_r$. Under the separability assumption (see \eqref{dual-rate}), $R_r = f_r^{-1} (\sum_{l':l' \in r}\lambda_{l'})$, so $y_l  = \sum_{r:l \in r} f_r^{-1} (\sum_{l':l' \in r} \lambda_{l'}).$
Let $\vec{y}$ be the vector with elements $y_l$ so that 
\begin{align}\label{eq:vecy}
\vec{y} = {\bf R}f^{-1}({\bf R}^T \vec{\lambda}),
\end{align}
where ${\bf R}$ is the routing matrix as defined in Section \ref{subsec:stability}. 
Here, the inverse function $f^{-1}(\cdot)$ is applied to individual elements of the corresponding input vector. 
Let $J_{\vec{y}}(\vec{\lambda})$ and $J_{\vec{f}}(\vec{R})$ denote the Jacobians of $\vec{y}$ and $\vec{f}$, where $\vec{y}$ and $\vec{f}$ are vector functions of $\vec{\lambda}$ and $\vec{R}$ respectively. I.e., $\vec{f}(\vec{R}) = [f_1(R_1), f_2(R_2),...f_r(R_r),...]$ (and likewise for $\vec{y}$). From \eqref{eq:vecy} and by applying the inverse function theorem (Theorem 2-11, \cite{spivak1965calculus}), the Jacobian matrix of $\vec{y}$ is given by:
\begin{align}\label{eq:Jy}
J_{\vec{y}}(\vec{\lambda}) = {\bf R}[J_{\vec{f}}(\vec{R})]^{-1}{\bf R}^T.
\end{align} 
\noindent Here, $[J_{\vec{f}}(\vec{R})]^{-1}$ is the matrix inverse of the Jacobian matrix $J_{\vec{f}}(\vec{R})$. We then compute
\begin{align}\label{eq:dot-vlambda}
&\sum_{l\in L}  \dot{V}_{\lambda_l} = \sum_{l\in L} (\lambda_l-\lambda_l^*)\dot{\lambda_l}  \nonumber\\
&=\bigg[\sum_{l\in L} (\lambda_l-\lambda_l^*)(y_l-y_l^*)\bigg] + (\vec{w}-\vec{w}^*)^T(\vec{\lambda} - \vec{\lambda}^*).
\end{align}

Concavity of $U$ guarantees that the diagonal elements of $J_{\vec{f}}(\vec{R})$ are at most zero. Hence, from \eqref{eq:Jy}, the diagonal elements of $J_{\vec{y}}(\vec{\lambda})$ are non-positive. Thus, the first term of the right-hand side of the \eqref{eq:dot-vlambda} is at most zero.  Combining these derivatives yields
\begin{eqnarray}  
\lefteqn{\dot{V} = \sum_{l\in L} \dot{V}_{w_l}  + \sum_{r\in {\cal R}} \dot{V}_{\mu_r} + \sum_{l\in L} \dot{V}_{\lambda_l}} \nonumber \\
&=& \sum_{l\in L}(w_l-w_l^*)(U'_{w_l}(w) - U'_{w_l}(w^*)) \nonumber \\
& & \mbox{} + \sum_{l\in L} \sum_{r: l \in r} [\mu_r((w_l - w_l^*)/w_l + \log(w_l^*/w_l)) \nonumber \\
& & \mbox{} +  \mu_r^*((w_l^* - w_l)/w_l^* + \log(w_l/w_l^*))] \nonumber \\
& & \mbox{} + (\vec{\lambda} - \vec{\lambda^*})^T (\vec{y} - \vec{y^*}) \leq 0.
\label{eq:overall Lyap derivative}
\end{eqnarray}
Since {\bf R} has full column rank, then $\vec{y} = \vec{y^*}$ implies $\vec{\lambda} = \vec{\lambda^*}$. Consequently,
$\dot{V} = 0$ on the set $M=\{(\vec{w}, \vec{\mu}, \vec{\lambda}): \vec{w}=\vec{w}^*, \vec{\lambda} = \vec{\lambda}^*\}.$

By LaSalle's invariance principle,  $(\vec{w}(t), \vec{\mu} (t), \vec{\lambda}(t))$ converges to the largest invariant set in $M$ which we claim to be the equilibrium point.  To see this,  suppose $(\vec{w}, \vec{\mu}, \vec{\lambda}) \in M$ with $\vec{\mu} \neq \vec{\mu}^*$. Note that, at equilibrium, $\dot{w_l} = 0,$ i.e.,
\begin{align}
    U'_{w_l}(\vec{w}^*) - \lambda_l^* + \sum_{r:l \in r} \mu_r^*/w_l^* = 0,
\end{align}
Thus, we have 
\begin{align}\label{eq:dotw_leq}
\dot{w_l} = \sum_{r:l \in r} (\mu_r - \mu_r^*)/w_l^*, \forall l.
\end{align}

Let $W$ be the diagonal matrix whose $(l, l)$th entry is $w_l^*$ with all off-diagonal elements equal to zero. Then, from \eqref{eq:dotw_leq}, 
$\dot{\vec{w}} = W^{-1}{\bf R}(\vec{\mu} - \vec{\mu^*}) \ne \vec{0}$, since {\bf R} has full column rank.
So, immediately, $(\vec{w}, \vec{\mu}, \vec{\lambda}) \notin M$.  Hence, $(w^*, \mu^*, \lambda^*)$ is the largest invariant set in $M$ and the equilibrium point is globally asymptotically stable.\\

\noindent
{\bf Non-Concave Utility -- local asymptotic stability}:  
With $\vec{x}=[\vec{w},\, \vec{\mu},\, \vec{\lambda}]^T$, let $\vec{\delta x}$ denote $\vec{x}$ perturbed from equilibrium $\vec{x}^*$.  The linearization of (\ref{eq: DE}) around $\vec{x}^*$ is $\dot{\vec{\delta x}} = A_{x^*} \vec{\delta x}$, where
\begin{eqnarray*}
A_{x^*} = \left[
\begin{array}{ccc}
B & W^{-1} {\bf R} & -I\\
-W^{-1} {\bf R} & 0 & 0\\
I & 0 & J_{\vec{y}}
\end{array}
\right],
\end{eqnarray*}
and where $B$ is a diagonal matrix with $(l, l)$th entry
\begin{equation}
U''_{w_l}(w_l^*) - \sum_{r:l \in r} \mu_r^*/{w_{l}^*}^2;
\label{eq:beta}
\end{equation}
$W$, $J_{\vec{y}}$ and ${\bf R}$ be the matrices defined as before. Now consider the Lyapunov candidate $V(\vec{\delta x}) = \vec{\delta x}^T \vec{\delta x}$ and form 
$$
A_{x^*} + A_{x^*}^T = 
\left[
\begin{array}{ccc}
B & 0 & 0\\
0 & 0 & 0\\
0 & 0 & J_{\vec{y}}
\end{array}
\right].
$$
If (\ref{eq:beta}) is negative for each $l$, then
$$
\dot{V}(\vec{\delta x}) = \vec{\delta x}^T(A_{x^*} + A_{x^*}^T)\vec{\delta x} \leq 0,
$$
for all $\vec{\delta x}$ and equal to zero when $\vec{\delta x} = [0,\, \vec{\delta \mu},\, 0]$. Using LaSalle's principle as in the concave utility case, we can conclude that $V$ is a Lyapunov function and equilibrium $\vec{x}^*$ is locally asymptotically stable.

\subsection{Handling q-datagram losses}\label{sub:qdatagram-loss-appendix}
Even with reliable classical communication and deterministic swaps, a q-datagram might still be lost if nodes have limited quantum memory. A node is considered \textit{congested} when all of its memory slots are occupied by q-datagrams awaiting swaps. In this case, admitting a new q-datagram requires overwriting an existing memory slot, which results in the loss of the previously-stored q-datagram. In simulations, we choose to overwrite the oldest qubit in memory when a node is congested. In the event of q-datagram losses while executing \dqnum, the state maintained by link controllers may become inconsistent. Given a session $r$ and a specific link $l$ along its  path, we define:
\begin{itemize}
    \item \emph{Upstream links} as those links that precede link $l$ along the path from the source to the sink of session $r$;
    \item \emph{Downstream links} as those links that follow link $l$ (including link $l$) along the same path.
\end{itemize}
If a q-datagram is lost just before reaching a particular link controller (say, at link $l$), upstream controllers will have applied state updates (e.g., to $R_l^{\text{sum}}$ and $M_l^{\text{sum}}$) while downstream controllers remain unaware of the dropped q-datagram: the internal states of upstream controllers is thus irreversibly out of date. To address this, we incorporate a \emph{back-propagation correction mechanism}. A classical control message is sent upstream, carrying the necessary values from the lost q-datagram (specifically, $\Delta R_r$ and $\Delta \mu_r$), 
so that each upstream controller can revert its local state by performing corrective updates to restore $R_l^{\text{sum}}$ and $M_l^{\text{sum}}$ to their previous values. I.e., for each upstream link $l_u$, we do the following updates:
\begin{equation}
    R_{l_u}^{\text{sum}} \leftarrow R_{l_u}^{\text{sum}} - \Delta R_r, \quad 
    M_{l_u}^{\text{sum}} \leftarrow M_{l_u}^{\text{sum}} - \Delta \mu_r.
\end{equation}

\begin{figure}[tb]
\centering
\includegraphics[width=\linewidth]{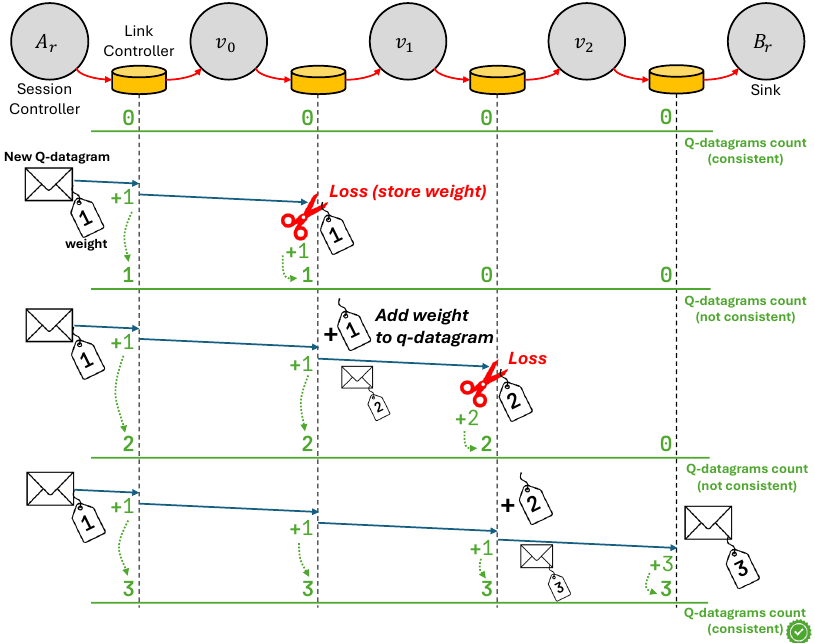}
\caption{Example sequence of the \textit{weight} mechanism to handle losses in \dqnumapproax. The example shows a situation where two consecutive q-datagrams are dropped (by different link controllers). The signed integers and the q-datagram counts, marked in green, track how many q-datagrams in total every link controller has processed at a given time. The first q-datagram that is successfully delivered to the sink has accumulated the weight ($3$ in the example) of all the previously lost q-datagrams. All link controllers have thus processed the same number ($3$) of q-datagrams and their rate estimates will be consistent.}
\label{fig:example_weights}
\end{figure}

In \dqnumapproax, the $\Delta R_r$ field is absent from the q-datagram header. Nevertheless, it is still essential to account for q-datagram losses, because downstream link controllers must maintain an accurate estimate of the aggregate rate $\sum_r R_r$. If losses are ignored, the aggregate rate will be systematically underestimated by downstream link controllers. To address this, we introduce an additional field in every q-datagram, called the \textit{weight}, which is initialized to $1$ when the q-datagram is created.
This field indicates how many q-datagrams the current q-datagram effectively represents. When link controllers process q-datagrams, they no longer simply count each as a single unit. Instead,  a q-datagram with weight $m$ contributes $m$ units to the aggregate rate. With the exponential average implementation of the rate estimate, the link controller applies \eqref{eq:approx-O} $m$ times, effectively treating the q-datagram as a bulk arrival of $m$ q-datagrams. When a link controller drops a q-datagram with weight $m$, it interprets this event as the loss of $m$ q-datagrams and stores this information internally. To ensure this loss is accounted for in downstream link controllers, the controller then increases the weight of the next successfully forwarded q-datagram from the same session by $m$. In this way, the weight field allows the system to propagate accurate rate information even in the presence of losses: every surviving q-datagram is effectively the sum of itself and all the q-datagrams that were lost before it along the path. We show an illustrative example in Fig. \ref{fig:example_weights}.

\end{document}